\documentclass[twocolumn,prd,aps,superscriptaddress,preprintnumbers,tightenlines,showpacs,nofootinbib,eqsecnum,amsfonts,amsmath]{revtex4-1}
\usepackage{epsfig}
\usepackage{graphics}
\usepackage{graphicx}
\usepackage{bm}
\usepackage[dvipsnames]{xcolor}
\usepackage{bm}
\usepackage{times}
\usepackage{xspace}
\usepackage[varg]{txfonts}
\usepackage[normalem]{ulem} 
\usepackage[colorlinks]{hyperref}
\usepackage[caption=false]{subfig}
\usepackage{booktabs}
\usepackage{url}
\usepackage{float}
\usepackage[bottom]{footmisc}
\usepackage{lineno}
\usepackage{makecell}

\definecolor{LinkColor}{rgb}{0.75, 0, 0}
\definecolor{CiteColor}{rgb}{0, 0.5, 0.5}
\definecolor{UrlColor}{rgb}{0, 0, 0.75}
\hypersetup{linkcolor=LinkColor}
\hypersetup{citecolor=CiteColor}
\hypersetup{urlcolor=UrlColor}
\usepackage{perpage}
\MakePerPage{footnote}

\newcommand{\paperone}{Paper~I\xspace}

\newcommand{\blambda}{\bm{\lambda}}

\newcommand{\bxigr}{\bm{\xi}_{\text{GR}}}
\newcommand{\bxingr}{\bm{\xi}_{\text{nGR}}}

\newcommand{\bs}[1]{\bm{\vec{S}_{#1}}}
\newcommand{\Mo}{M_{\odot}}

\newcommand{\df}[1]{\delta f_{\text{#1}}}
\newcommand{\dtau}[1]{\delta \tau_{\text{#1}}}
\newcommand{\fngr}[1]{f_{\text{#1}}}
\newcommand{\taungr}[1]{\tau_{\text{#1}}}
\newcommand{\fgr}[1]{f ^{\text{GR}}_{\text{#1}}}
\newcommand{\taugr}[1]{\tau ^{\text{GR}}_{\text{#1}}}
\newcommand{\pSEOB}{\texttt{pSEOBNR}}
\newcommand{\SEOB}{\texttt{SEOBNR}}

\newcommand{\AEI}{\affiliation{Max Planck Institute for Gravitational Physics (Albert Einstein Institute), Am M\"uhlenberg 1, Potsdam 14476, Germany}}
\newcommand{\UMD}{\affiliation{Department of Physics, University of Maryland, College Park, MD 20742, USA}}

\begin{document}

\title{Constraints on quasi--normal-mode frequencies with LIGO-Virgo binary--black-hole observations} 

\author{Abhirup Ghosh}
\AEI
\author{Richard Brito}
\affiliation{Dipartimento di Fisica, ``Sapienza" Universit\`a di Roma $\&$ Sezione INFN Roma1, Piazzale Aldo Moro 5, 00185 Roma, Italia}
\author{Alessandra Buonanno}
\AEI
\UMD

\date{\today}

\begin{abstract}
  The no-hair conjecture in General Relativity (GR) states that the
  properties of an astrophysical Kerr black hole (BH) are completely described by its
  mass and spin. As a consequence, the complex
  quasi-normal-mode (QNM) frequencies of a binary--black-hole (BBH)
  ringdown can be uniquely determined by the mass and spin of the
  remnant object. Conversely, measurement of the QNM frequencies could
  be an independent test of the no-hair conjecture. This paper extends to spinning BHs earlier work that proposed to
  test the no-hair conjecture by measuring the complex QNM
  frequencies of a BBH ringdown using parameterized inspiral-merger-ringdown waveforms in the effective-one-body formalism, 
thereby taking full advantage of the entire signal power and removing dependency on the
  predicted or estimated start time of the ringdown. Our method was used to analyze the
  properties of the merger remnants for BBHs observed by
  LIGO-Virgo in the first half of their third observing (O3a) run.
 After testing our method with GR and non-GR synthetic-signal injections in Gaussian noise, 
we analyze, for the first time, two BBHs from the first (O1) and second (O2) LIGO-Virgo 
  observing runs, and two additional BBHs from the O3a run. We then provide joint constraints with published results
  from the O3a run. In the most agnostic and conservative scenario where we combine the information from different
events using a hierarchical approach, we obtain, at $90\%$ credibility, that 
the fractional deviations in the frequency (damping time) of the dominant QNM are 
$\df{220}=0.03^{+0.10}_{-0.09}$ ($\dtau{220}=0.10^{+0.44}_{-0.39}$), respectively, an improvement of a factor of $\sim 4$ ($\sim 2$) 
over the results obtained with our model in the LIGO-Virgo publication. The single-event most-stringent constraint to date continues to be 
GW150914 for which we obtain $\df{220}=0.05^{+0.11}_{-0.07}$ and $\dtau{220}=0.07^{+0.26}_{-0.23}$.
\end{abstract}

\maketitle

\section{Introduction}
\label{sec:intro}
The LIGO Scientific Collaboration~\cite{lsc} and the Virgo
Collaboration~\cite{Virgo} have recently announced their catalogue of
gravitational-wave (GW) signals from the first
half of the third observing run (O3a)~\cite{Abbott:2020niy}. Combined
with the first and second observing-run catalogues~\cite{LIGOScientific:2018mvr}, the Advanced LIGO detectors at Hanford,
Washington and Livingston, Louisiana~\cite{TheLIGOScientific:2014jea},
and the Advanced Virgo detector in Cascina,
Italy~\cite{TheVirgo:2014hva} have now detected $50$ GW
events from the merger of compact objects like neutron stars and/or
black holes (BHs). Alongside independent claims of
detections~\cite{Nitz:2018imz,Nitz:2019hdf,Venumadhav:2019lyq,Zackay:2019btq}, these results have firmly established the field of GW astronomy, five years after the first detection of a GW passing through Earth, GW150914~\cite{Abbott:2016blz}.

The observation of GWs has had significant astrophysical and cosmological
implications~\cite{TheLIGOScientific:2016htt,GBM:2017lvd,Monitor:2017mdv,Abbott:2017xzu}. It
has also allowed us to probe fundamental
physics and test predictions of Einstein's theory of General Relativity
(GR) in the previously unexplored highly-dynamical, and strong field
regime~\cite{TheLIGOScientific:2016src,Abbott:2018lct,LIGOScientific:2019fpa,Abbott:2020jks}. In GR, a binary black hole (BBH) system is described by
three distinct phases: an early \textit{inspiral}~\cite{Blanchet:2013haa}, where the two
compact objects spiral in losing energy because of the emission of GWs, a \textit{merger}~\cite{Pretorius:2005gq,Campanelli:2005dd,Baker:2005vv} marked by the
formation of a common apparent horizon, and a \textit{ringdown}~\cite{Vishveshwara:1970zz,Vishveshwara:1970cc,Press:1971wr,Chandrasekhar:1975zza,Detweiler:1980gk}, during which the newly formed remnant object settles down to a Kerr BH emitting quasi-normal-modes (QNMs) (i.e., damped oscillations with specific, discrete frequencies and decay times). 

The no-hair conjecture in GR~\cite{Israel:1967wq,Hawking:1971vc,Carter:1971zc,Robinson:1975bv,Mazur:1982db} states that an (electrically neutral) astrophysical BH is completely described by two observables: mass and spin. One
consequence of the no-hair conjecture is that the (complex) QNM
frequencies of gravitational radiation emitted by a perturbed isolated
BH are uniquely determined by its mass and spin. Hence
a test of the no-hair conjecture would involve checking for
consistency between estimates of mass and spin of the remnant object
across multiple QNM frequencies~\cite{Dreyer:2003bv,Berti:2005ys}.  Consistency of the late-time waveform with a single QNM is a test of the ringdown of
a BBH coalescence, but not necessarily a test of the no-hair
conjecture, which requires the measurement of (at least) two QNMs (BH
spectroscopy), and consistency between them~\cite{Dreyer:2003bv,Gossan:2011ha, Meidam:2014jpa,Carullo:2018sfu,Carullo:2018gah,Isi:2019aib,Bhagwat:2019bwv,Cabero:2019zyt,Maselli:2019mjd}. An inconsistency would either
indicate a non-BH nature of the remnant object, or an incompleteness
of GR as the underlying theory of gravity.

The LIGO-Virgo collaborations have released companion papers detailing their results
of tests of GR for GW150914 \cite{TheLIGOScientific:2016src}, and for several GW events of
the two transient catalogues (TC): GWTC-1 \cite{LIGOScientific:2018mvr,LIGOScientific:2019fpa}  and GWTC-2 \cite{Abbott:2020niy,Abbott:2020jks}.
The results include tests of GW generation and source dynamics, where bounds are placed on
parameterized deviations in the post-Newtonian (PN) coefficients describing
the early inspiral, and
phenomenological coefficients describing the intermediate (plunge) and
merger regimes of coalescence~\cite{Arun:2006hn,Arun:2006yw,Agathos:2013upa,Abbott:2018lct}; tests of GW
propagation, which assume a generalized dispersion relation and place
upper bounds on the Compton wavelength and, consequently, the mass of
the graviton~\cite{Abbott:2017vtc,Samajdar:2017mka}, and tests of the
polarization of gravitational radiation using a
multi--GW-detector network~\cite{Abbott:2017oio,Isi:2017fbj}. The GWTC-1/2 papers also
check for consistency between different portions of the signal using estimates for the predicted mass and spin of the remnant
object~\cite{Ghosh:2016xx,Ghosh:2017gfp,TheLIGOScientific:2016src}, and
consistency of the residuals with detector noise~\cite{Ghonge:2020suv,LIGOScientific:2019fpa}. None of these tests report any
departure from the predictions of GR.

The first paper on tests of GR by the LIGO Collaboration~\cite{TheLIGOScientific:2016src} also provided us with the first measurement of the dominant damped-oscillation signal in the ringdown stage of a BBH
coalescence, and more recently a similar measurement was made with the high-mass event GW190521~\cite{Abbott:2020tfl,Abbott:2020mjq}. The set of available measurements was greatly expanded in the latest LIGO-Virgo O3a testing GR paper~\cite{Abbott:2020jks} where a comprehensive analysis of the properties of the remnant object, including the ringdown stage, was reported for tens of GW events. The consistency between the post-merger signal and the least damped QNM was first demonstrated in
Ref.~\cite{TheLIGOScientific:2016src} for GW150914, and confirmed through several independent analyses~\cite{Brito:2018rfr,Carullo:2019flw,Isi:2019aib,CalderonBustillo:2020tjf}. This was later extended to include
overtones in Refs.~\cite{Giesler:2019uxc,Isi:2019aib,Abbott:2020jks}.  The nature of the remnant object has also been explored through tests of BH thermodynamics, including the Hawking's area
theorem~\cite{Cabero:2017avf,Isi:2020tac}, the Bekenstein-Hod universal bound~\cite{Carullo:2021yxh}, the BH area quantization~\cite{Foit:2016uxn,Laghi:2020rgl}, the consistency of the merger frequency with predictions from NR simulations~\cite{Carullo:2018gah} or through search for echos in the post-merger signal~\cite{Nielsen:2018lkf,Tsang:2019zra,Lo:2018sep,Abedi:2018npz,Abedi:2020sgg,Testa:2018bzd}.  None of these tests have found evidence for non-BH nature of the remnant
object (as described in GR) in LIGO-Virgo BBH observations.

With current ground-based detectors, only a small fraction of BBH coalescences lead to a detectable post-merger GW signal and, even for these events, the signal-to-noise ratio (SNR) in the post-merger signal is, in most cases, only slightly above threshold for detection~\cite{LIGOScientific:2019fpa,Abbott:2020jks}.
This is particularly important, given that most of the tests mentioned above restrict the data to this post-merger signal which, while being agnostic to deviations in the pre-merger dynamics, significantly reduces parameter-estimation capabilities. Added to a lack of SNR is the ambiguity in clearly defining a start time for the ringdown phase (discussed, for example, in Refs.~\cite{Berti:2007fi,Baibhav:2017jhs,Bhagwat:2017tkm}). Defining such a time is particularly important in order to avoid possible systematic errors that may occur if one tries to extract the fundamental QNMs too early after merger, since the higher-order overtones dominate the ringdown signal closer to merger~\cite{Buonanno:2006ui,Giesler:2019uxc}. In some tests~\cite{Carullo:2018gah,Carullo:2019flw}, the ringdown start time has been left as a free parameter to be estimated directly from the data. Whereas, in other cases, the ringdown start time is predicted from corresponding full-signal parameter-estimation analyses (see \texttt{pyRing} analysis in Ref.~\cite{Abbott:2020jks}). Uncertainties
in estimates of the ringdown start-time, as well as an overall lack of
SNR in the post-merger signal, given typical sensitivities of
ground-based detectors, can result in significant statistical uncertainties in the measurement of some of the QNM properties. This is especially true for the measurement of the QNM damping times which are, in general, harder to measure from the data. Hence, one might want to look at alternate methods to measure QNMs using information from as much as the signal as possible.

An independent approach to BH spectroscopy, based on the full-signal
analysis, was introduced in Ref.~\cite{Brito:2018rfr}
(henceforth referred to as \paperone). Unlike methods that focus only
on the post-merger signal, it employs the
complete inspiral-merger-ringdown
(IMR) waveform to measure the complex QNM
frequencies. In the current version, this method assumes that GR provides a very good description of the signal up to merger, while being agnostic about the complex QNM frequencies. This allows to access the full SNR of the signal, reducing
measurement uncertainties. However, we note that this method can incorporate deviations from GR, notably from the (known) PN coefficients,  
also during the long inspiral stage~\cite{Abbott:2018lct,
LIGOScientific:2019fpa,Abbott:2020jks}, and can in principle incorporate deviations from GR, notably from the parameters calibrated to NR simulations,  
in the late inspiral and plunge. Moreover, in this method, the definition of the
ringdown start time is built into the merger-ringdown model and does not need to be
either left as an additional free parameter or fixed using alternate
definitions. While \paperone presented the method and tested it for non-spinning BBHs, here we extend the analysis to the more
realistic astrophysical case in which BHs carry spin. Furthermore,
the IMR waveform model used in this paper is more accurate than what employed in \paperone,
because it contains higher-order corrections in PN theory and it was calibrated to a much larger set
of numerical-relativity (NR) waveforms~\cite{Bohe:2016gbl}. All astrophysical BHs are expected to be spinning,
and ignoring effects of spin has been shown to introduce systematic biases in the measurement of the
source properties.

The rest of the paper is organized as follows. Section~\ref{sec:model} describes our parameterized IMR
waveform model. In Sec.~\ref{sec:method}, we define our framework to test GR, notably how we can measure
complex QNM frequencies with our parameterized model within a Bayesian formalism, and validate it using 
several synthetic-signal injection studies in Gaussian noise in Sec.~\ref{sec:results}. Then, in Sec.~\ref{sec:results_lvc},
we apply our method on real GW events in LIGO and Virgo data. In particular, we analyze
the GWTC-2 events and obtain the most stringent constraints on the dominant (complex) QNM frequecies. Finally, in Sec.~\ref{sec:discussion} we provide a summary of
our results and discuss future developments.


\section{Waveform model and statistical strategy to measure quasi-normal modes}
A GW signal from the (quasi-circular) coalescence of two BHs is
completely described in GR by $15$ parameters,
$\bxigr$. These can be grouped into the \emph{intrinsic} parameters:
the (source) masses, $m_1, m_2$ and spins, $\bs1, \bs2$ of the component
objects in the binary; and the
\emph{extrinsic} parameters: a reference time $t_c$ and phase
$\phi_c$, the sky position of the binary ($\alpha$,
$\delta$), the luminosity distance, $d_L$, and the binary's orientation
described through the inclination of the binary $\iota$ and its
polarization $\psi$. We also introduce the total (source) mass $M = m_1+m_2$,
and the (dimensionless) symmetric mass ratio $\nu = m_1m_2/M^2$. We also follow the convention $m_1 > m_2$, and hence the asymmetric mass ratio, $q = m_1/m_2 \geq1$. We also introduce the detector masses $m_{1\,\rm det}= (1+z)\,m_1$ and $m_{2 \,\rm det} = (1+z)\,m_2$, 
where $z$ is the redshift.

Here, we focus on BHs with spins aligned or anti-aligned
with the orbital angular momentum (henceforth, aligned-spin). In this case,
the GW signal depends on $11$ parameters. We denote the
aligned-spin (dimensionless) components as $\chi_{i} = |\vec{\bm{S}}_i|/m^2_i$, where $i=1,2$ for the two BHs.

\subsection{Parameterized waveform model}\label{sec:model}

As in \paperone, we use an IMR waveform model developed within the effective-one-body (EOB)
formalism~\cite{Buonanno:1998gg,Buonanno:2000ef}. However, whereas \paperone was limited to non-spinning multipolar waveforms,
here we use as our baseline model the aligned-spin multipolar waveform model
developed in Ref.~\cite{Cotesta:2018fcv}. In addition to being
calibrated to NR simulations, this model also uses information from BH
perturbation theory for the merger and ringdown phases. Henceforth we
will denote this model by $\SEOB$ for short~\footnote{In the LIGO Algorithm Library (LAL), this
waveform model is called {\tt SEOBNRv4HM}.}.

In the observer's frame, the GW polarizations can be written as
\begin{equation}
h_+(\iota,\varphi_0;t ) - i h_\times(\iota,\varphi_0;t) = \sum_{\ell, m} {}_{-\!2}Y_{\ell m}(\iota,\varphi_0)\, h_{\ell m}(t)\,,
\end{equation}
where $\varphi_0$ is the azimuthal direction to the observer (note that without loss of generality we can take $\phi_c\equiv\varphi_0$), while ${}_{-\!2}Y_{\ell m}(\iota,\varphi_0)$ are the $-2$ spin-weighted spherical harmonics where $(\ell,m)$ are the usual indices that describe the angular dependence of the spin-weighted spherical harmonics, with $\ell\geq 2$, $-\ell\leq m\leq \ell$. The $\SEOB$ model we employ includes the $(\ell, |m|)=(2,2),(2,1)$, $(3,3)$, $(4,4)$, and $(5,5)$ modes~\cite{Cotesta:2018fcv}. For each $(\ell, m)$, the inspiral-(plunge-)merger-ringdown $\SEOB$ waveform is schematically given by
\begin{equation}
h_{\ell m}(t) = h_{\ell m}^\mathrm{insp-plunge}\, \theta(t_\mathrm{match}^{\ell m} - t) + h_{\ell m}^\mathrm{merger-RD}\,\theta(t-t_\mathrm{match}^{\ell m})\,,
\end{equation}
where $\theta(t)$ is the Heaviside step function, $h_{\ell m}^\mathrm{insp-plunge}$ represents the inspiral-plunge part of the waveform, whereas $h_{\ell m}^\mathrm{merger-RD}$ denotes the merger-ringdown waveform, which reads~\citep{Bohe:2016gbl,Cotesta:2018fcv}
\begin{equation}
\label{RD}
h_{\ell m}^{\textrm{merger-RD}}(t) = \nu \ \tilde{A}_{\ell m}(t)\ e^{i \tilde{\phi}_{\ell m}(t)} \ e^{-i \sigma_{\ell m 0}(t-t_{\textrm{match}}^{\ell m})},
\end{equation}
where $\nu$ is the symmetric mass ratio of the binary and $\sigma_{\ell m0} = 2\pi f_{\ell m 0} -i/\tau_{\ell m 0}$ denotes the complex frequency of the fundamental QNMs of the remnant BH, i.e. QNMs with overtone index $n=0$. We denote the oscillation frequencies by $f_{\ell m  0}\equiv \Re(\sigma_{\ell m0})/(2\pi)$ and the decay times by $\tau_{\ell m 0}\equiv -1/\Im(\sigma_{\ell m0}) $.
The functions $\tilde{A}_{\ell m}(t)$ and $\tilde{\phi}_{\ell m}(t)$ are given by~\cite{Bohe:2016gbl,Cotesta:2018fcv}:
\begin{subequations}
\begin{eqnarray}
\label{eq:ansatz_amp}
\tilde{A}_{\ell m}(t) &=& c_{1,c}^{\ell m} \tanh[c_{1,f}^{\ell m}\ (t-t_{\textrm{match}}^{\ell m}) \ +\ c_{2,f}^{\ell m}] \ + \ c_{2,c}^{\ell m},\\
\label{eq:ansatz_phase}
\tilde{\phi}_{\ell m}(t) &=& \phi_{\textrm{match}}^{\ell m} - d_{1,c}^{\ell m} \log\left[\frac{1+d_{2,f}^{\ell m} e^{-d_{1,f}^{\ell m}(t-t_{\textrm{match}}^{\ell m})}}{1+d_{2,f}^{\ell m}}\right],
\end{eqnarray}
\end{subequations}
where $ \phi_{\textrm{match}}^{\ell m}$ is the phase of the inspiral-plunge mode $(\ell, m)$ computed at $t = t_{\textrm{match}}^{\ell m}$. The coefficients $d_{1,c}^{\ell m}$ and $c_{i,c}^{\ell m}$ with $i = 1,2$
are fixed by imposing that the functions $\tilde{A}_{\ell m}(t)$ and $\tilde{\phi}_{\ell m}(t)$ are of class $C^1$ at $t = t_{\textrm{match}}^{\ell m}$, when matching the merger-ringdown waveform to the inspiral-plunge $\SEOB$ waveform $h_{\ell m}^\mathrm{inspiral-plunge}(t)$. This allows us to write the coefficients $c_{i,c}^{\ell m}$ as~\cite{Cotesta:2018fcv}:
\begin{subequations}
\begin{align}
\label{c1}
c_{1,c}^{\ell m} &= \frac{1}{c_{1,f}^{\ell
    m} \nu} \big[ \partial_t|h_{\ell
    m}^{\textrm{insp-plunge}}(t_{\textrm{match}}^{\ell m})| \nonumber \\
    &- \sigma^\textrm{R}_{\ell m} |h_{\ell
    m}^{\textrm{insp-plunge}}(t_{\textrm{match}}^{\ell
    m})|\big] \cosh^2{(c_{2,f}^{\ell m})}, \\
\label{c2}
c_{2,c}^{\ell m} &= -\frac{ |h_{\ell
    m}^{\textrm{insp-plunge}}(t_{\textrm{match}}^{\ell
    m})|}{\nu} + \frac{1}{c_{1,f}^{\ell
    m} \nu} \big[ \partial_t|h_{\ell
    m}^{\textrm{insp-plunge}}(t_{\textrm{match}}^{\ell m})|  \nonumber \\
    &- \sigma^\textrm{R}_{\ell m} |h_{\ell
    m}^{\textrm{insp-plunge}}(t_{\textrm{match}}^{\ell
    m})|\big] \cosh{(c_{2,f}^{\ell m})}\sinh{(c_{2,f}^{\ell m})}, \\ \nonumber
\end{align}
\end{subequations}
and $d_{1,c}^{\ell m}$ as
\begin{align}
\label{d1}
d_{1,c}^{\ell m} &= \left[\omega_{\ell m}^{\textrm{insp-plunge}}(t_{\textrm{match}}^{\ell m}) -  \sigma^\textrm{I}_{\ell
      m}\right]\frac{1+ d_{2,f}^{\ell m}}{d_{1,f}^{\ell m}d_{2,f}^{\ell m}}\,,
\end{align}
where we denoted $\sigma_{\ell m}^\textrm{R} \equiv \Im (\sigma_{\ell m0}) < 0$ and  $\sigma_{\ell m}^\textrm{I} \equiv -\Re (\sigma_{\ell m0})$, and $\omega_{\ell m}^{\textrm{insp-plunge}}(t)$ is the frequency of the inspiral-plunge EOB mode. The coefficients $c_{i,f}^{\ell m}$ and $d_{i,f}^{\ell m}$ are obtained through fits to NR and
Teukolsky-equation--based waveforms and can be found in Appendix C of Ref.~\cite{Cotesta:2018fcv}.

In the $\SEOB$ model constructed in Ref.~\cite{Cotesta:2018fcv}, the
complex frequencies $\sigma_{\ell m 0}$ are expressed in terms of the
final BH mass and spin~\cite{Berti:2005ys,Berti:2009kk}, and the
latter are related to the BBH's component masses and spins through
NR--fitting-formulas obtained in
GR~\cite{Taracchini:2013rva,Hofmann:2016yih}. Here instead, in the
spirit of what was done in \paperone, we promote the QNM (complex)
frequencies to be free parameters of the model, while keeping the
inspiral-plunge modes $h_{\ell m}^\mathrm{inspiral-plunge}(t)$ fixed
to their GR values. More explicitly, we introduce a parameterized
version of the $\SEOB$ model where the frequency and the
damping time of the ${\ell m 0}$ mode (i.e, $(f_{\ell m 0}, \tau
_{\ell m 0})$) are defined through the fractional deviations, $(\delta
f_{\ell m 0},\delta \tau_{\ell m 0})$, from the corresponding GR
values~\cite{Gossan:2011ha,Meidam:2014jpa}.

Thus,
\begin{subequations}
\begin{eqnarray}
f_{\ell m 0} &=& f_{\ell m 0}^{\text{GR}}\, (1 + \delta f_{\ell m 0})\,,\label{eq:nongr_freqs_a} \\ 
\tau _{\ell m 0} &=& \tau _{\ell m 0}^{\text{GR}}\, (1 + \delta \tau_{\ell m 0})\,. \label{eq:nongr_freqs_b}
\end{eqnarray}
\end{subequations}

The GR quantities $( f_{\ell m 0}^{GR},\tau_{\ell m 0}^{GR})$ are
constructed using the NR--fitting--formula from Refs.~\cite{Taracchini:2013rva,Hofmann:2016yih}, and are functions of the initial masses and spins, $(m_1, m_2, \chi_1, \chi_2)$. Hence,
\begin{subequations}
\begin{eqnarray}
f_{\ell m 0} &=& f_{\ell m 0}(m_1, m_2, \chi_1, \chi_2, \delta f_{\ell m 0}, \delta \tau_{\ell m 0})\,,\\ 
\tau _{\ell m 0} &=& \tau _{\ell m 0}(m_1, m_2, \chi_1, \chi_2, \delta f_{\ell m 0}, \delta \tau_{\ell m 0})\,.
\end{eqnarray}
\end{subequations}
We denote such a parameterized waveform model $\pSEOB$~\footnote{This
waveform model is called {\tt pSEOBNRv4HM} in LAL.}. We note that when leaving $\sigma_{\ell m}$ to vary
freely, the functions $\tilde{A}_{\ell m}(t)$ and $\tilde{\phi}_{\ell
  m}(t)$ in general also differ from the GR predictions, since
those functions depend on the QNM complex frequencies, as can be seen
from the expressions for $c_{i,c}^{\ell m}$ and $d_{1,c}^{\ell m}$ in Eqs.~(\ref{c1}),
(\ref{c2}), and (\ref{d1}). As a consequence, the ringdown signal (amplitude and phase) 
soon after merger deviates from the one predicted by GR.

\subsection{Bayesian parameter-estimation technique}
\label{sec:method}

The parameterized model, $\pSEOB$, described above introduces an additional set of non-GR parameters, $\bxingr = (\delta f_{\ell m 0},\delta \tau_{\ell m 0})$, corresponding to each $(\ell,m)$ QNM present in the GR waveform model $\SEOB$. One then proceeds to use the Bayes' theorem to obtain the \emph{posterior} probability distribution on $\blambda = \{\bxigr, \bxingr\}$, given a hypothesis $\mathcal{H}$:
\begin{equation}
P(\blambda | d, \mathcal{H}) = \frac{P(\blambda | \mathcal{H}) \, \mathcal{L}(d | \blambda, \mathcal{H})}{P(d|\mathcal{H})},
\label{eq:Bayes_theorem}
\end{equation}
where $P(\blambda | \mathcal{H})$ is the \emph{prior} probability distribution, and $\mathcal{L}(d | \blambda, \mathcal{H})$ is called the \emph{likelihood} function. The denominator is a normalization constant $P(d|\mathcal{H}) = \int P(\blambda | \mathcal{H}) \, \mathcal{L}(d | \blambda, \mathcal{H}) \, d\blambda$, called the marginal likelihood, or the \emph{evidence} of the hypothesis $\mathcal{H}$. In this case, our hypothesis $\mathcal{H}$ is that the data contains a GW signal that is described by the $\pSEOB$ waveform model $h(\blambda)$  and stationary Gaussian noise described by a power spectral density (PSD) $S_n(f)$. The likelihood function can consequently be defined as:
\begin{equation}
\mathcal{L}(d | \blambda, \mathcal{H}) \propto \exp\big[-\frac{1}{2} \langle d - h(\blambda) \, | \, d -h(\blambda) \rangle \big],
\label{eq:likelihood}
\end{equation}
where $\langle . | . \rangle$ is the usual noise-weighted inner product:
\begin{equation}
\langle A | B \rangle = \int_{f_\mathrm{low}} ^{f_\mathrm{high}} df \frac{\tilde{A}^*(f)\tilde{B}(f) + \tilde{A}(f)\tilde{B}^*(f)}{S_n(f)}.
\label{eq:nwip}
\end{equation}
The quantity $\tilde{A}(f)$ denotes the Fourier transform of $A(t)$ and the $^*$ indicates complex conjugation. The limits of integration ${f_\mathrm{low}}$ and ${f_\mathrm{high}}$ define the bandwidth of the sensitivity of the GW detector. We usually assume ${f_\mathrm{high}}$ to be the Nyquist frequency whereas ${f_\mathrm{low}}$ is dictated by the performance of the
GW detector at low-frequency. Here, we follow the choice made in the LIGO-Virgo analysis~\cite{LIGOScientific:2018mvr,Abbott:2020niy}. Namely, for all events and injections we consider, we set ${f_\mathrm{low}}=20$ Hz, except for the GW190521-like injection and the real GW190521 event for which use ${f_\mathrm{low}}=11$ Hz~\cite{Abbott:2020tfl,Abbott:2020mjq}. Owing to the large dimensionality of the parameter set $\blambda$, the posterior distribution $P(\blambda | d, \mathcal{H})$ in Eq.~(\ref{eq:Bayes_theorem}) is computed by stochastically sampling the parameter space using techniques such as Markov-chain Monte Carlo (MCMC)~\cite{Metropolis:1953am,Hastings:1970aa} or Nested Sampling~\cite{Skilling:2006gxv}. For this paper, we use the \verb+LALInference+~\cite{Veitch:2014wba} and \verb+Bilby+ codes~\cite{Ashton:2018jfp,Smith:2019ucc,Speagle_2020} that provide an implementation of the parallely tempered MCMC and Nested Sampling algorithms respectively, for computing the posterior distributions.

Given the full-dimensional posterior probability density function $P(\blambda | d, \mathcal{H})$, we can marginalize over the \emph{nuisance} parameters, to obtain the marginalized posterior probability density function over the QNM parameters $\bxingr$:

\begin{equation}
P(\bxingr | d, \mathcal{H})= \int P(\blambda | d, \mathcal{H}) d\bxigr\,.
\end{equation}

For most of the results discussed in this paper, we restrict ourselves
to the $(\ell m) = (2,2)$ and/or $(3,3)$ modes. In those cases we assume $\bxingr = \{\df{220},\dtau{220}\}$ and/or $
\{\df{330},\dtau{330}\}$, and fix all the other $(\ell m)$ modes to their GR predictions (i.e., $\delta f_{\ell m 0} = \delta
\tau_{\ell m 0} = 0$). This is because, for most of the high-mass BH
events that we find most appropriate for this test, the LIGO-Virgo
observations are consistent with nearly--equal-mass face-on/off BBHs
for which power in the subdominant modes is not enough to
attempt to measure more than one QNM complex frequency.

Lastly, throughout our analysis, we assume uniform priors on our non-GR QNM
parameters, $(\delta f_{\ell m 0},\delta \tau_{\ell m 0})$. We note that
since the priors on $( f_{\ell m 0}^{GR},\tau_{\ell m 0}^{GR})$ are
derived through NR--fits, from the corresponding priors on the initial
masses and spins, this leads to a non-trivial prior on the final
reconstructed frequency and damping time, $( f_{\ell m 0},\tau_{\ell m
  0})$. Also, given the definition of the damping time in
Sec.~\ref{sec:model}, we note that $\delta \tau_{\ell m 0} = -1$ leads
to the imaginary part of the QNM complex frequency going to infinity. We avoid
this by restricting the minimum of the prior on $\delta \tau_{\ell m
  0}$ to be greater than $-1$.


\section{Synthetic-signal injection study}
\label{sec:results}
\subsection{Simulations using GR signals in Gaussian noise} \label{ssec:gr_signal}

We now demonstrate our method using synthetic-signal injections describing GWs
from BBHs in GR. We employ coloured Gaussian noise with PSDs for LIGO and
Virgo detectors during the fourth observing (O4) run~\cite{Abbott:2020qfu}, which 
is expected to start in the second half of 2022~\cite{AdvLIGOPSD,TheVirgo:2014hva}.
For the mock BBH signals, we choose parameters similar to two specific GW events, GW150914~\cite{Abbott:2016blz} and
GW190521~\cite{Abbott:2020tfl}. We list them in Table~\ref{tab:injection_values}.
These two binary systems are representative of the kind of systems for which
the QNM measurement is most suitable, notably high-mass BBH events which are loud enough that the
pre- and post-merger SNRs return reliable parameter-estimation results.

\begin{table}[h!]
\begin{center}
\begin{tabular}{ c|c|c|c|c|c|c|c|c }

Injection &  Network & \makecell{$m_{\rm 1\,det}$ \\$(\Mo)$} &  \makecell{$m_{\rm 2\,det}$ \\ $(\Mo)$} & $\chi_{1}$ & $\chi_{2}$ & $\rho_\text{IMR}$ & $\rho_\text{insp}$ & $\rho_\text{postinsp}$ \\
 \hline
 GW150914-like & HL &39 & 31 & 0.0 & 0.0 & 25 & 22 & 12 \\
 GW190521-like & HL & 150 & 120 & 0.02 & -0.39 & 20 & 8 & 18 \\
 SXS:BBH:0166 & HLV &72 & 12  & 0.0 & 0.0 & 71 & 58 & 41 \\

\end{tabular}
\caption{Parameters of the synthetic-signal injections, chosen to be similar to the actual GW events indicated in the first column (first two rows). The parameters $(m_{\rm 1 \,det}, m_{\rm 2 \,det})$ are the detector-frame masses of the primary and secondary BHs, respectively. The third row indicates the parameters of the SXS BBH waveform used in Sec.~\ref{ssec:nohairtheorem}. The second column refers to the detector-network used, with H,L,V, referring to LIGO-Hanford, LIGO-Livingston and Virgo, respectively. The quantities $\rho_\text{IMR}$, $\rho_\text{insp}$ and $\rho_\text{postinsp}$ are the SNR of the full IMR signal, SNR upto a certain cutoff frequency, and SNR after the cutoff frequency respectively. In each case, the cutoff frequency is assumed to be the frequency at the innermost circular stable orbit (ISCO) corresponding to the remnant Kerr BH.}
\label{tab:injection_values}
\end{center}
\end{table}

\begin{figure*}[hbt]
\begin{center}
        \includegraphics[width=0.5\textwidth]{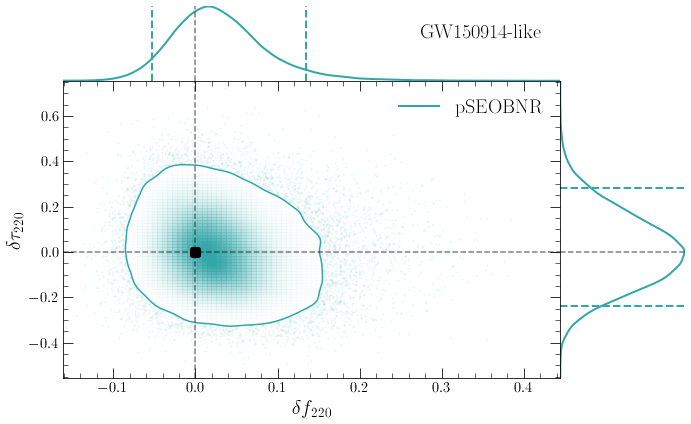}\includegraphics[width=0.5\textwidth]{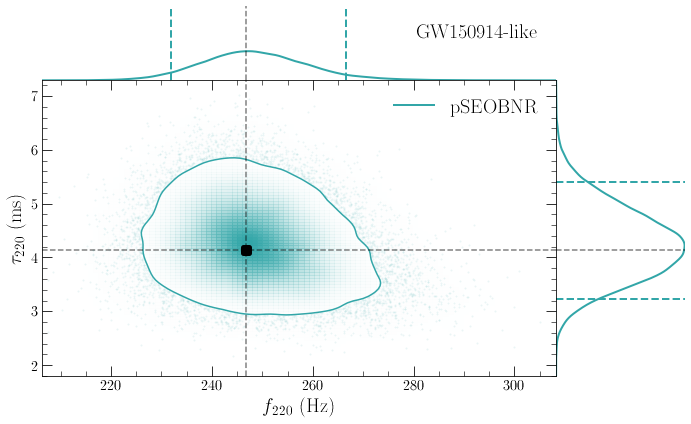}
        \includegraphics[width=0.5\textwidth]{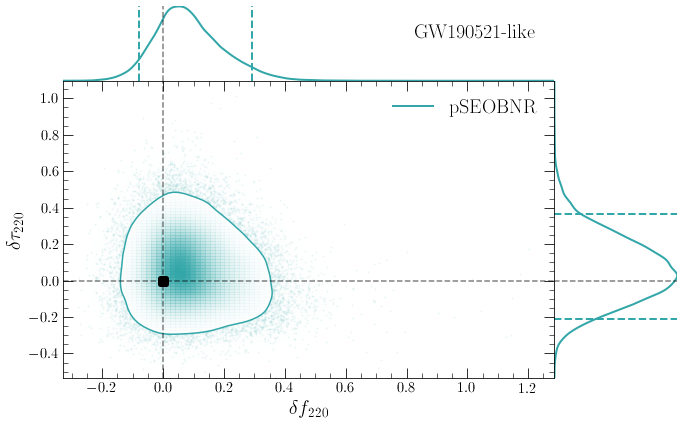}\includegraphics[width=0.5\textwidth]{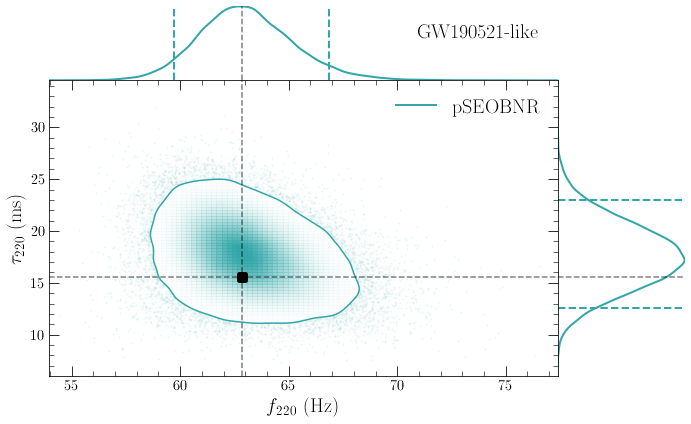}
        \caption{Posterior probability
          distribution on the fractional deviations in the frequency
          and damping time of the $(2,2)$ QNM, $(\df{220},\dtau{220})$
          (left panels) and the reconstructed quantities,
          $(\fngr{220}, \taungr{220})$ (right panels) for GR
          injections ($\SEOB$) with initial parameters similar to GW150914 (top
          panels) and GW190521 (bottom panels)
          (see Table~\ref{tab:injection_values}). The 2D contour marks the
          90\% credible region, while the dashed lines on the 1D
          marginalized distributions mark the 90\% credible
          levels. The black vertical and horizontal lines mark the
          injection values.}
        \label{fig:simulated_signal_GR}
\end{center}
\end{figure*}

To avoid possible systematic biases in our parameter-estimation analysis
due to error in waveform modeling, we use the GR version of the same waveform,
$\SEOB$ (i.e., without allowing for deviations in the QNM parameters) to
simulate our GW signal. And to avoid systematic biases due to noise,
we use an averaged (zero-noise) realization of the noise~\footnote{A detailed
study on noise systematics for one of the GW events is presented in
Appendix~\ref{sec:noise_systematics}.}.  Since mearly--equal-mass binaries like GW150914 and
GW190521 observed at moderately high SNRs are not expected to have a
loud ringdown signal, we restrict ourselves to estimating the
frequency and damping time of only one QNM $(\ell m) = (2,2)$ (i.e.,
$\{\df{220},\dtau{220}\}$), while fixing the other QNM frequencies to
their GR values.

We find, as one might expect, that the posterior distribution on the
parameters describing fractional deviations in the frequency and
damping time are consistent with zero (left panels of
Fig.~\ref{fig:simulated_signal_GR}). One can then convert these
fractional quantities into absolute quantities using the relations
given in Eqs.~(\ref{eq:nongr_freqs_a}) and ~(\ref{eq:nongr_freqs_b}), and
construct posterior distributions on these effective quantities,
$(\fngr{220}, \taungr{220})$ (right panels of
Fig.~\ref{fig:simulated_signal_GR}). In each of these cases, the recovered
two-dimensional posteriors are consistent with the GR predictions
(black dashed lines).

\subsection{Simulations using non-GR signals in Gaussian noise} \label{ssec:ngr_signal}

\begin{figure}
        \includegraphics[width=0.5\textwidth]{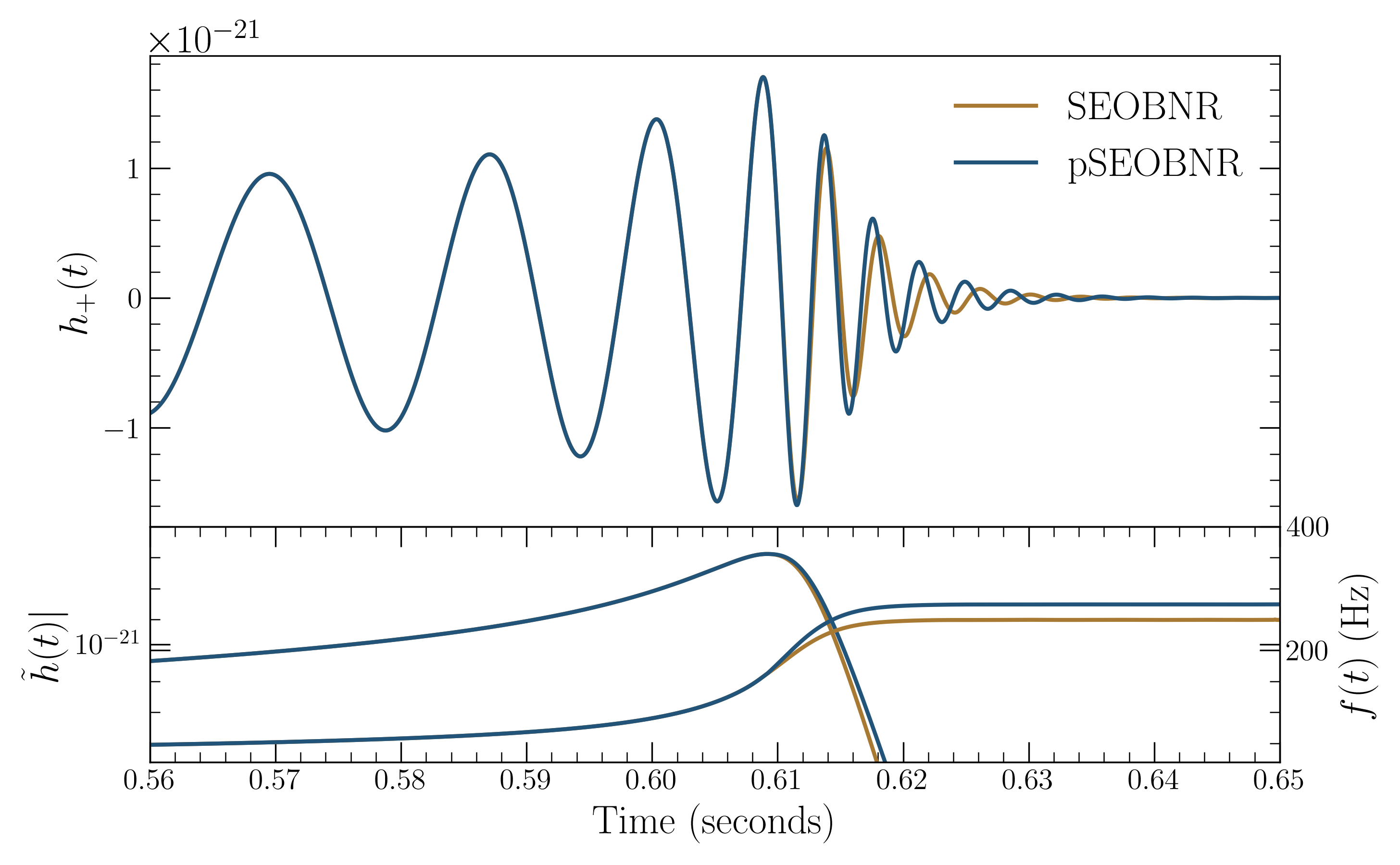}
        \caption{Top panel: The `+'--polarization of the gravitational waveform $h_+(t)$ from a GW150914-like event where the post-merger is described by GR (i.e., $\df{220} = \dtau{220} = 0$), and where the merger-ringdown is modified (i.e., $\df{220} = \dtau{220} = 0.1$). Bottom panel: Comparison of the evolution of the amplitude, $\tilde{h}(t)$ (left) and instantaneous frequency, $f(t)$ (right) for the GR and non-GR signal.}
        \label{fig:nongr_waveform}
\end{figure}

\begin{figure*}
\begin{center}
        \includegraphics[width=0.5\textwidth]{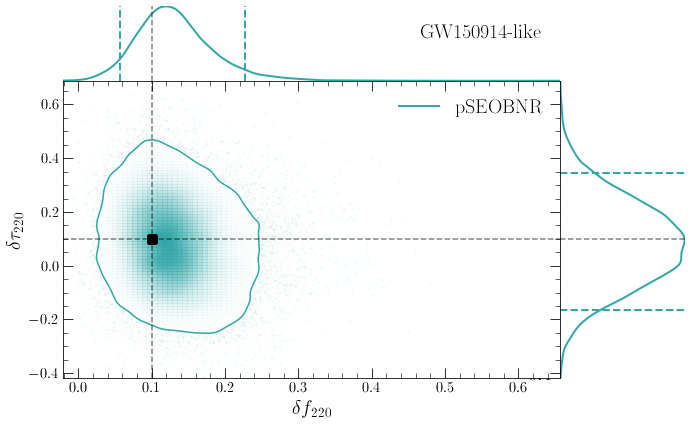}\includegraphics[width=0.5\textwidth]{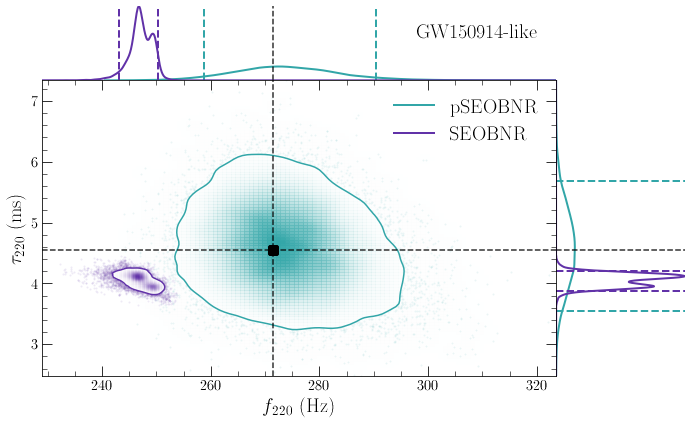}
        \includegraphics[width=0.5\textwidth]{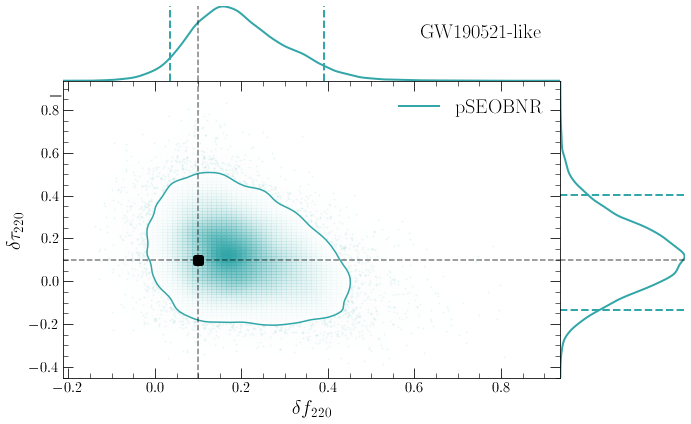}\includegraphics[width=0.5\textwidth]{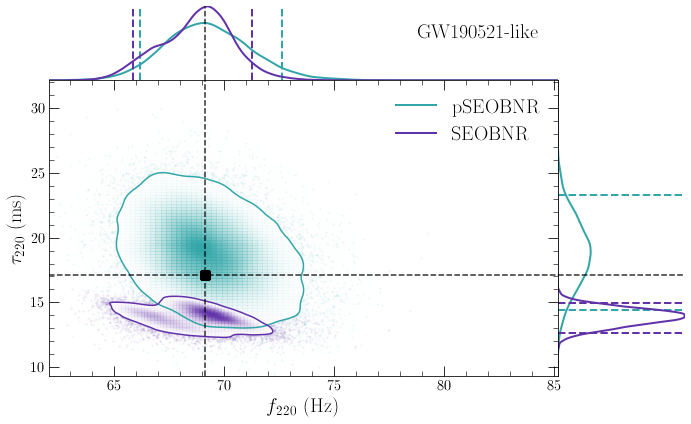}
        \caption{Posterior probability distribution on the fractional deviations in the frequency and damping time of the $(2,2)$ QNM, $(\df{220},\dtau{220})$ (left panels) and the reconstructed quantities, $(\fngr{220}, \taungr{220})$ (right panels) for non-GR injections ($\pSEOB$) with parameters of GW150914-like (top panels) and GW190521-like (bottom panels) as given in Table~\ref{tab:injection_values}. The non-GR signal has a deviation, $\df{220} = \dtau{220} = 0.1$. The 2D contour marks the 90\% credible region, while the dashed lines on the 1D marginalized distributions mark the 90\% credible levels. The black vertical and horizontal lines mark the injection values. In the right panels, we additionally show measurements using a GR ($\SEOB$) waveform, for the
GW150914-like (upper panel) and GW190521-like (lower panel) injections. The measurements with $\SEOB$ waveforms are visibly biased.}
        \label{fig:simulated_signal_nonGR}
\end{center}
\end{figure*}

To demonstrate the robustness of the method in detecting possible
deviations from GR, we inject synthetic GW signals which are identical
to the corresponding GR prediction up to merger, and differ in their
post-merger description. We again choose binary-parameters similar to
GW150914 and GW190521 (see Table ~\ref{tab:injection_values}), but set
$\df{220} = \dtau{220} = 0.1 $.  In other words, we assume that the
frequency and damping time of our non-GR signal is 10\% more than the
corresponding GR prediction, although the pre-merger signal is
identical to GR. In Fig.~\ref{fig:nongr_waveform} we show this non-GR
waveform, \texttt{pSEOBNR} with respect to the original GR template,
\texttt{SEOBNR}. We see that the waveforms are identical in amplitude
and instanteneous frequency upto the merger (lower panel) , beyond
which the red (GR template) and blue (non-GR template) differ. We
summarize the results of the Bayesian analysis in
Fig.~\ref{fig:simulated_signal_nonGR} where we show the posterior
probability distributions for $(\df{220}, \dtau{220})$, or
equivalently $(\fngr{220}, \taungr{220})$. We find that they are
consistent with the corresponding values of the injection parameters,
indicated by the black dashed lines.  We also note that at the
  SNRs we consider, since statistical uncertainties dominate
  systematic biases, the measurement excludes, at the 90\% credible
  level, the GR prediction of the frequency ($\df{220}=0$), but
  includes the prediction of the damping time
  ($\dtau{220}=0$). However, with louder events, one would expect
  these measurement errors to shrink and the $\taungr{220}$
  measurement to be inconsistent with the GR prediction, as well.

\begin{figure*}
        \includegraphics[width=0.5\textwidth]{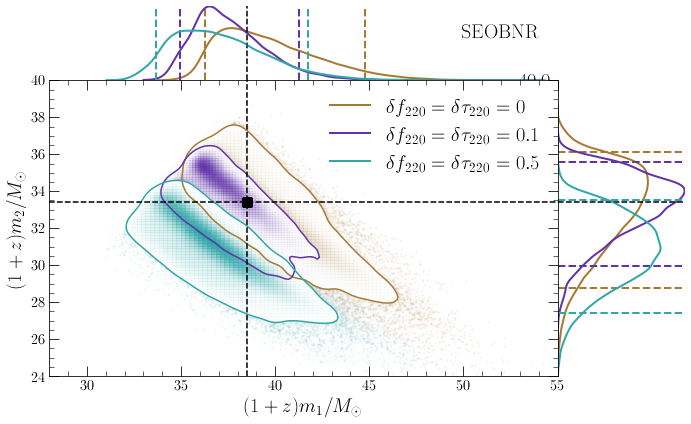}\includegraphics[width=0.5\textwidth]{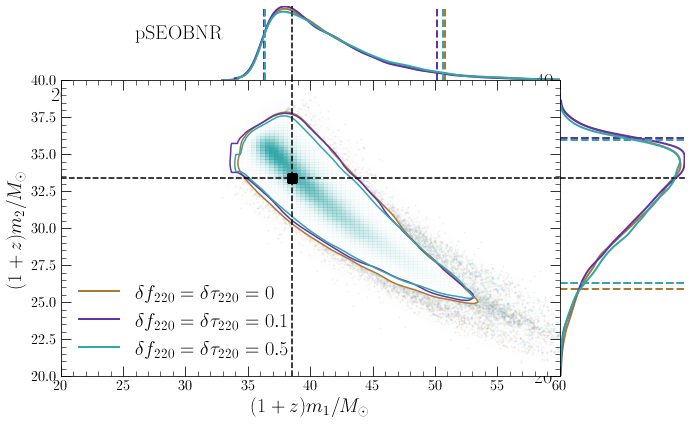}
        \includegraphics[width=0.5\textwidth]{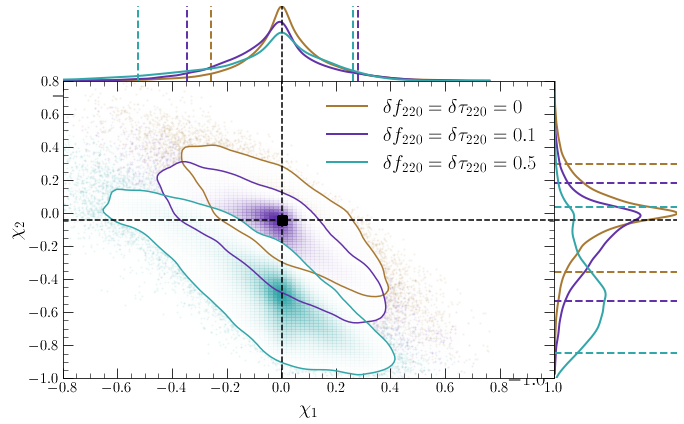}\includegraphics[width=0.5\textwidth]{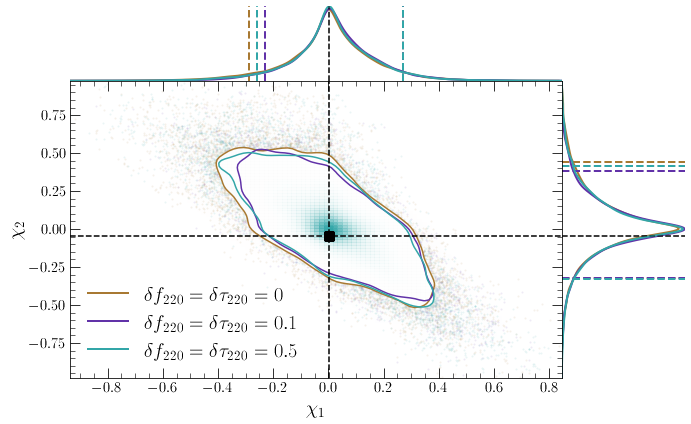}
        \includegraphics[width=0.5\textwidth]{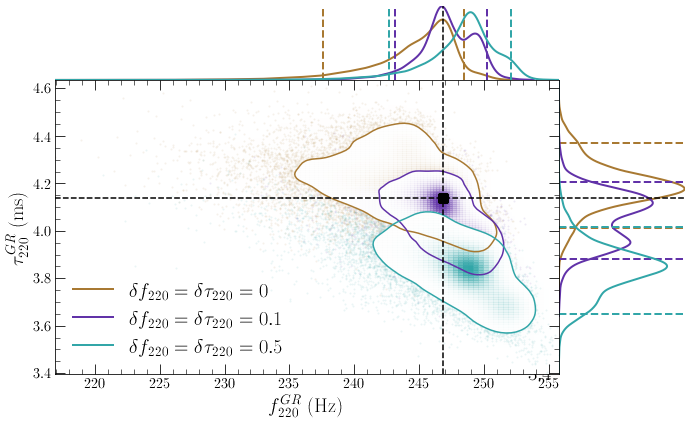}\includegraphics[width=0.5\textwidth]{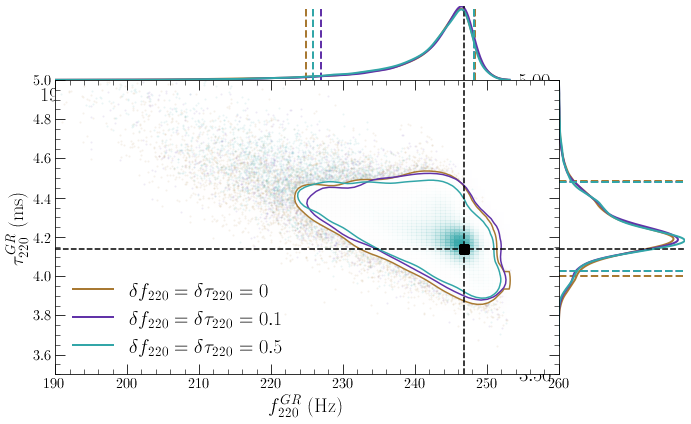}
        \includegraphics[width=0.5\textwidth]{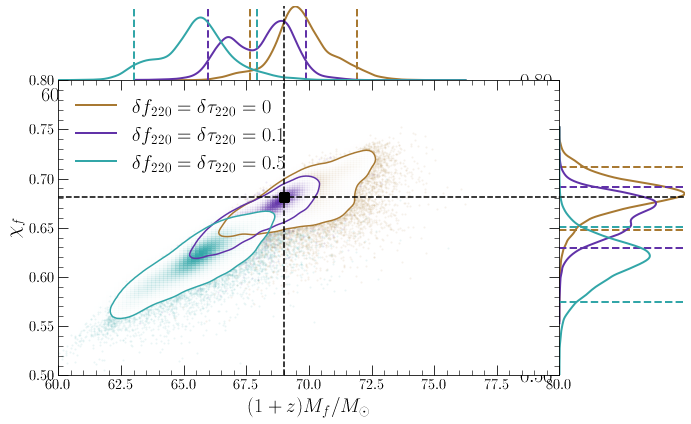}\includegraphics[width=0.5\textwidth]{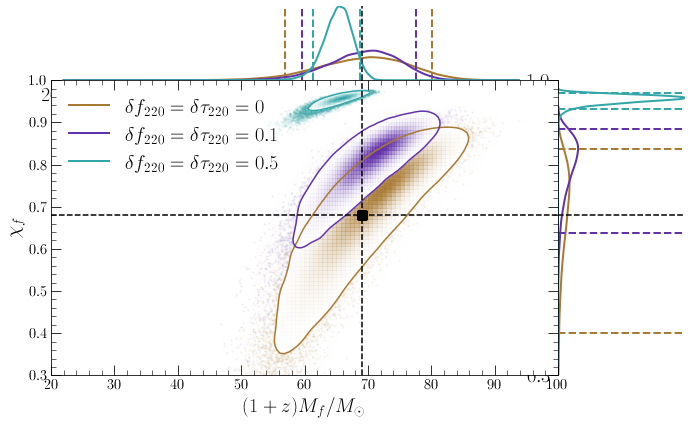}
         \caption{Comparison of a GW150914-like injection's parameters when signals with zero, 10\% and 50\% deviations ($\df{220} = \dtau{220} = 0, 0.1, 0.5$) are recovered using a GR ($\SEOB$) (left panels) or a non-GR ($\pSEOB$) (right panels) waveform model. The panels (from top to bottom) show the 2D posteriors (with 90\% credible levels) and corresponding marginalised 1D posteriors (with 90\% credible levels) in (detector-frame) masses (first row), dimensionless spins (second row), GR predictions of frequency and damping time (third row) and the remnant mass and spin predictions ($M_f$, $\chi_f$) (fourth row) from the frequency and damping time. In each case, the injection values are indicated by the black dashed lines. In the lowermost panel, the injection values of the final mass and spin, correspond to the injection with no deviations. The $\df{220} = \dtau{220} = 0.1$ signal is identical to the results shown in Fig.~\ref{fig:simulated_signal_nonGR}.}
        \label{fig:gr_ngr_comparison}
\end{figure*}

We additionally investigate the effects of erroneously assuming that
an underlying non-GR signal can be well-described by a GR one. We do
this by estimating the parameters of our non-GR signals using the GR
waveform model $\SEOB$ instead of the parameterized $\pSEOB$.  The
resulting one- and two-dimensional posteriors are shown in the right
panels of Fig.~\ref{fig:simulated_signal_nonGR} by purple curves for the
GW150914-like (top) and GW190521-like (bottom) signals
respectively. For both signals, we find the 2D $\SEOB$ estimates are
markedly biased with respect to the $\pSEOB$ estimates.  We investigate 
this impact on the estimation of the GR parameters $\bxigr$ by changing the
magnitude of deviation in the non-GR parameters more in Fig.~\ref{fig:gr_ngr_comparison}. We restrict ourselves 
to a GW150914-like event, and for comparison, add a synthetic signal
with $\df{220} = \dtau{220} = 0.5$, or 50 \% deviation, alongside the
10\% non-GR signal and a signal with no deviation (essentially GR)
mentioned above. The left (right) panels show the
posterior probability distributions of the three signals using the GR,
$\SEOB$ (non-GR, $\pSEOB$) waveform model. The first difference we
note is the $\SEOB$ recoveries yield biased estimates when the
underlying signal is non-GR, while the $\pSEOB$ recoveries do
not. Furthermore, as we increase the deviations, while the $\SEOB$
recoveries expectedly get more biased, the $\pSEOB$ measurements are
robust in consistently recovering the injected value. This gives us 
confidence that the $\pSEOB$ model can accurately measure QNM 
properties without biasing measurements of inspiral quantities, like masses and spins. 
As we increase the deviation, the remnant object rings down at a higher frequency and damping time. The resulting $\pSEOB$ signal is longer than the GR ($\SEOB$) prediction. When we try to fit the $\SEOB$ model to this signal (left column of Fig.~\ref{fig:gr_ngr_comparison}), the template tries to fit parameters appropriate for a longer signal, i.e., smaller masses and spins. Hence, with increasing deviations, the recovered masses and spins have a tendency to shift towards lower values.
To compute the samples of the final mass and spin in the last row, we start from the samples
of the complex QNM frequencies $(\fngr{220}, \taungr{220})$, which are obtained from the 
fractional deviation samples $(\df{220},\dtau{220})$ and the GR-quantities
$(\fgr{220},\taugr{220})$ using Eqs.~(\ref{eq:nongr_freqs_a}) and Eq.~(\ref{eq:nongr_freqs_b}),
and then invert them using the fitting formula in Ref.~\cite{Berti:2005ys}.
The three BBH signals, $\df{220} = \dtau{220} = 0, 0.1, 0.5$, correspond to
three unique sets of values for $(M_f,\chi_f)$. These predicted values are correctly 
recovered by the $\pSEOB$ waveform (bottom-right plot in Fig.~\ref{fig:gr_ngr_comparison}),
leading to three distinct and disjointed 2D posteriors on $(M_f,\chi_f)$, unlike the $\SEOB$
analysis (to reduce clutter in the plot, we just plot the injection values for $\df{220} = \dtau{220} = 0$).

\subsection{Test of the no-hair conjecture}\label{ssec:nohairtheorem}

Finally, we provide a simple demonstration of a test of the no-hair
theorem using our model. As described in the introduction, any test of
the no-hair theorem of BHs would need to involve independent
measurements of (at least) two different QNMs.

Here, we use an NR GW signal from the SXS catalog~\cite{Mroue:2013xna}
corresponding to a non-spinning BBH with mass-ratio $q=6$ (SXS:BBH:0166) and 
total mass $M=84 \Mo$ (see Table~\ref{tab:injection_values}).
We choose an asymmetric system to increase the SNR in the higher modes.
We also choose the distance and orientation of the binary
such that the total SNR in the three-detector network of LIGO Hanford, Livingston and
Virgo, is {$\sim$ 70}. Based on the LIGO-Virgo observations during the first three observing runs, 
such asymmetric and loud signals are no longer just a theoretical
prediction, but quite plausible during O4. Using this
signal, we attempt to measure both the $(2,2)$ and
$(3,3)$ QNMs. For this injected signal the SNR in
other sub-dominant modes is too low to be able to measure them.

\begin{figure}
        \includegraphics[width=0.5\textwidth]{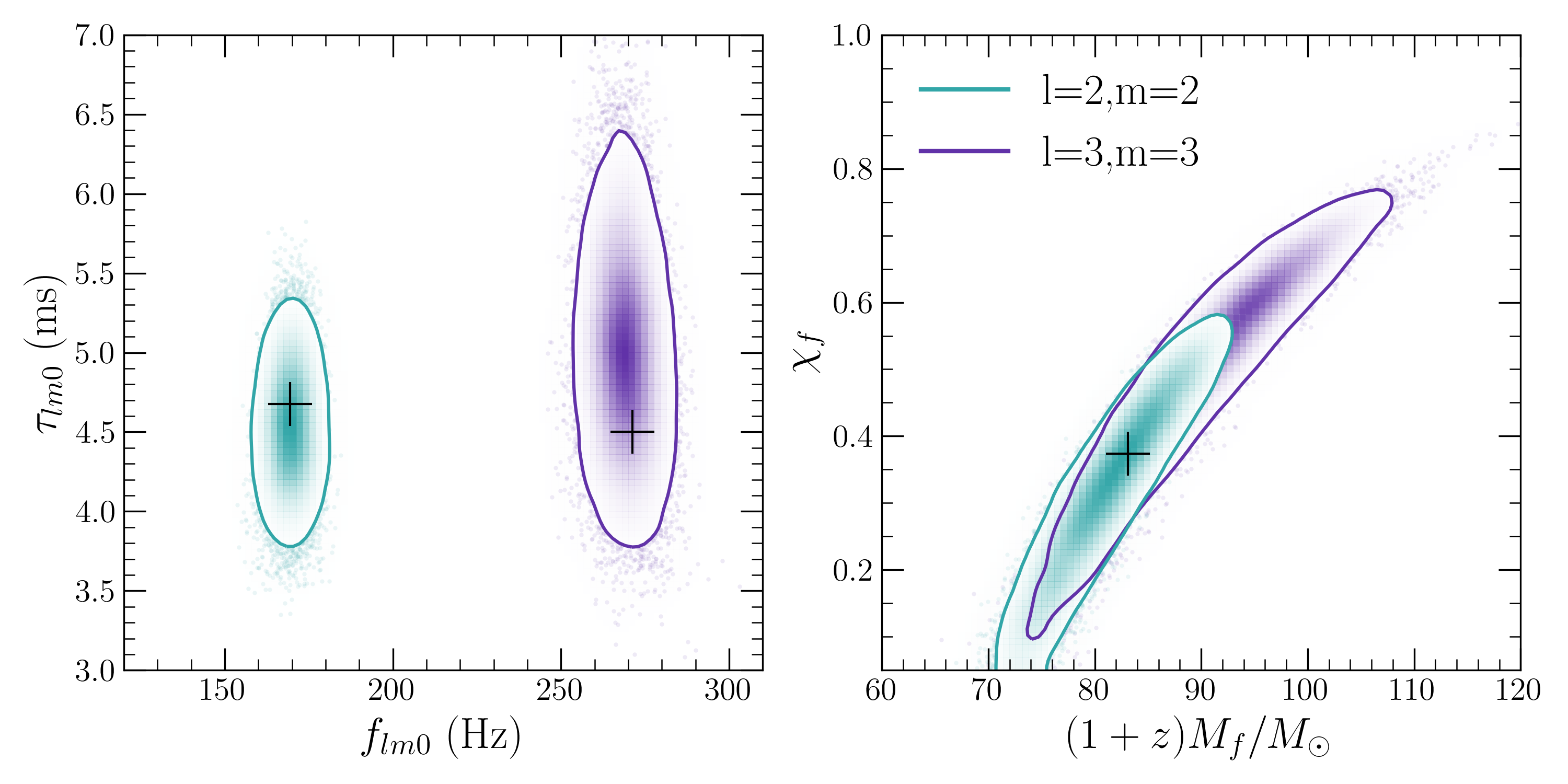}
        \caption{Posterior probability distribution on the frequency and damping time of the $(2, 2)$ and $(3, 3)$ QNM (left panel), and the final mass and spin inferred from the complex frequencies (right panel), when a NR signal with parameters $q=6$,  $M=84 \Mo$ and SNR $=75$ is injected in Gaussian noise and recovered with the $\pSEOB$ waveform model. The plus signs mark the GR predictions.}
        \label{fig:nohair_sxs}
\end{figure}

We summarize our results in Fig.~\ref{fig:nohair_sxs}.  Given the
injection parameters, the predicted values of the $(2, 2)$ and
$(3, 3)$ frequency and damping time are {(169.45 Hz, 4.68
  ms)} and {(271.21 Hz, 4.50 ms)} respectively. The left panel
of Fig.~\ref{fig:nohair_sxs} shows that the 2D posteriors
on the $(2, 2)$ and $(3, 3)$ QNMs are consistent with the
predictions for a BBH merger in GR, indicated by the black plus sign.
Using fitting formulae provided in Ref.~\cite{Berti:2005ys}, specifically,
Eq.~(2.1), Eq.~(E1) and Eq.~(E3) and Tables VIII and IX for the fitting coefficients,
we infer the 2D posterior probability distribution on the
final mass and spin as predicted independently by the $(2, 2)$ (blue) and $(3, 3)$ (red)
QNMs in the right panel of Fig.~\ref{fig:nohair_sxs}. The two
independent estimates are consistent with each other and correspond to
a unique mass and spin for the remnant BH {(83.08 $\Mo$, 0.37)}
indicated by the plus sign. As a consequence, this may be considered
as a test of the no-hair conjecture. For most of the events observed
so far, the power in the $(3, 3)$ has not been sufficient to
measure it along with the $(2, 2)$. However, it might also be possible to combine information from
multiple observation over the coming few years to obtain meaningful
constraints on the $(3, 3)$ and other sub-dominant QNMs~\cite{Gossan:2011ha,Meidam:2014jpa,Yang:2017zxs}.


\section{Constraints on QNM frequencies using LIGO-Virgo data}
\label{sec:results_lvc}
\begin{figure*}
        \includegraphics[width=0.5\textwidth]{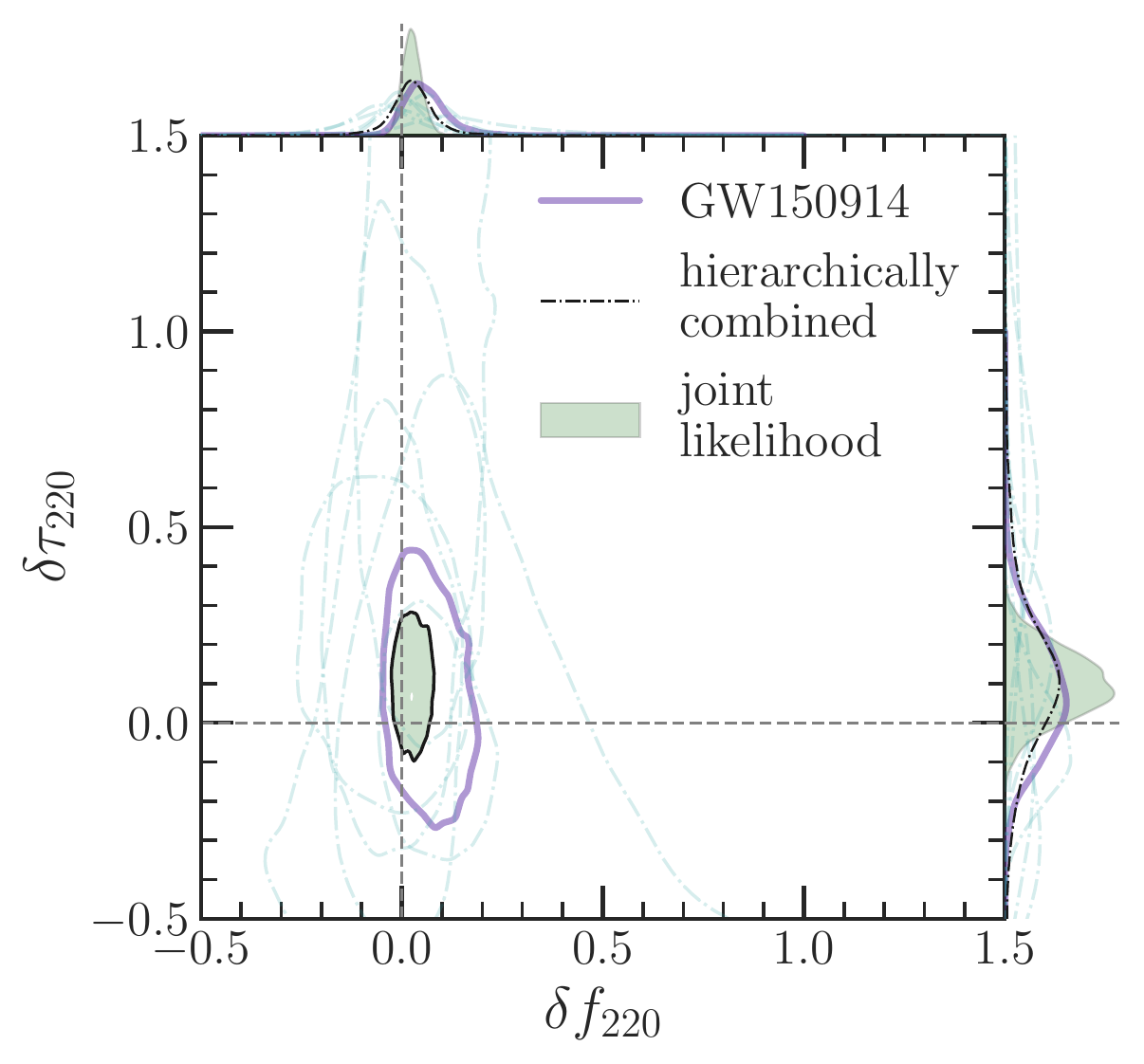}\includegraphics[width=0.3\textwidth]{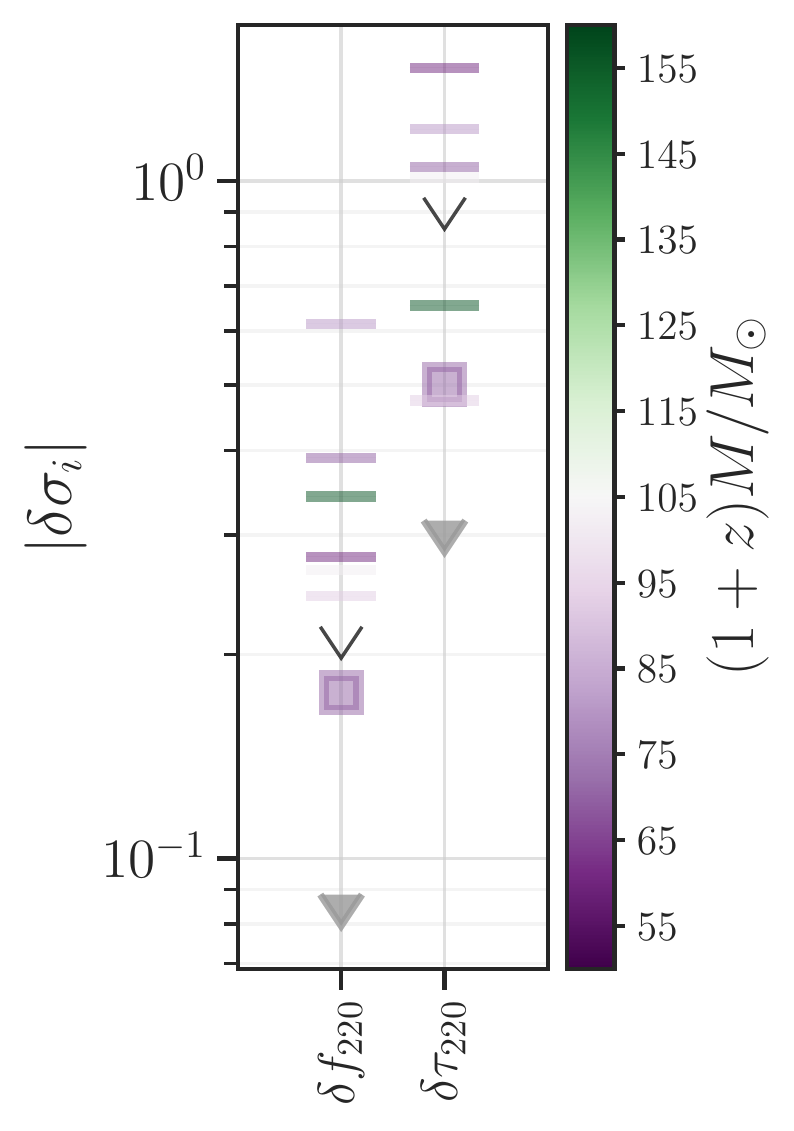}
        \caption{\emph{Left panel}:The 90\% credible levels of the posterior probability distribution of the fractional deviations in the frequency and damping time of the $(2, 2)$ QNM, $(\df{220},\dtau{220})$ and their corresponding one-dimensional marginalized posterior distributions, for events from O1, O2 and O3a passing a SNR threshold of $8$ in both the pre- and post-merger signal. The solid purple curve marks the best single-event constraint, GW150914, whereas the contraints from the other events are indicated by the dash-dot curves. The joint constraints on $(\df{220},\dtau{220})$ obtained multiplying the likelihoods from individual events is given by the filled grey contours, while the hierarchical method of combination yields the black dot dashed curves only shown in the 1D posteriors plots). \emph{Right panel}: 90\% credible interval on the one-dimensional marginalised posteriors on $\delta \sigma_i=(\df{220},\dtau{220})$, colored by the median redshifted total mass $(1 + z)M$, inferred assuming GR. Filled gray (unfilled black) triangles mark the constraints obtained when all the events are combined by multiplying likelihoods (hierarchically). The purple square marker indicates constraints from the single most-prominent event, GW150914.}
        \label{fig:o1o2_events}
\end{figure*}

The LIGO-Virgo Collaboration recently released their testing GR
catalogue containing results for events observed
during O3a~\cite{Abbott:2020jks}. For the test that we here present, the results shown in Ref.~\cite{Abbott:2020jks} only include events which pass a threshold for the median detector-frame total mass $\geq 90 \Mo$ and SNRs in the pre- and
post-merger regions $\geq 8$~\footnote{The pre- and post-merger regions of the signal are identified from the signal's power before and after the signal reaches the peak's amplitude, which is determined from the maximum of the likelihood function obtained with the parameter-estimation analysis. The SNR values are listed in Table IV of Ref.~\cite{Abbott:2020jks}}. 
The SNR threshold ensures that the signal contains sufficient
information in both the inspiral and merger stages to break the
degeneracy between the binary's total mass and the non-GR deviations
$(\df{220}, \dtau{220})$. Such strong degeneracy arises from the fact that the complex QNM frequencies are inversely proportional to the total mass~\cite{Berti:2005ys} such that one can always obtain the same fundamental frequency or damping time by appropriately increasing (or decreasing) the total mass and the deviation parameters in the same direction. This degeneracy is especially important in
low-SNR events with negligible higher-modes and for which only the
post-merger is detectable, rendering the measurement of $(\df{220},
\dtau{220})$ impossible for those cases  (see Appendix~\ref{sec:correlation} for more details). On the other hand the
total mass threshold $\geq 90 \Mo$ was employed due to the fact that
this analysis is computationally expensive, and also because we
expected these events to be the most promising for ringdown
studies. However, since the SNR threshold alone should be sufficient
for the analysis, for this paper we run the test on all the events
listed in Ref.~\cite{Abbott:2020jks} that have SNRs in the pre- and
post-merger regions $\geq 8$, without imposing any mass threshold.

Given the above we add the signals  GW190630$\_$185205 and GW190828$\_$063405, to the list of GW events considered in Ref.~\cite{Abbott:2020jks}. Furthermore, for the first time, 
we apply our method to measure the QNMs to GW events from O1 and O2, notably GW150914 and GW170104. The other high-mass events from O1 and O2, GW170729, GW170809,
GW170814, GW170818 and GW170823 do not have an SNR $\geq 8$ in the
merger-ringdown signal. The list of the signals for which we run the analysis is given in Table~\ref{tab:qnm_o1o2_results}.
  
For all the relevant signals, we show the posterior distributions
$(\df{220}, \dtau{220})$ in the left panel of
Fig.~\ref{fig:o1o2_events} and also provide the reconstructed QNM
parameters, $(\fngr{220}, \taungr{220})$ in
Table~\ref{tab:qnm_o1o2_results}. In the right panel of
Fig.~\ref{fig:o1o2_events} we also provide a summary of the 90\%
credible intervals on the 1D marginalized posteriors. We highlight the
dependence of the constraints on the total mass of the
system. In general, the tightest bounds are set by the 
most massive systems, as they tend to have larger post-merger SNR. We
find a similar trend in the right panel of Fig.~\ref{fig:o1o2_events}.
 
Among all the GW signals detected so far, GW150914 (solid curve in
Fig.~\ref{fig:o1o2_events}) is unique in its loudness, leading to the
first, and to date, best attempt in measuring the QNM
frequencies~\cite{TheLIGOScientific:2016src,Brito:2018rfr,Carullo:2019flw,Isi:2019aib}. Within
90\% credibility we obtain from GW150914:
\begin{equation}\label{GW150914_delta}
\df{220}=0.05^{+0.11}_{-0.07}\,\quad \dtau{220}=0.07^{+0.26}_{-0.23}\,.
\end{equation}

Stronger constraints can be obtained by combining information from all the events~\cite{Abbott:2020jks}. If we assume that the fractional deviations $(\df{220},\dtau{220})$ take the same values in multiple events, we can assume
the posterior of one event to be the prior for the next, and obtain a
joint posterior probability distribution. For $N$ observations, where
$P(\df{220}, \dtau{220} | d_j)$ is the posterior for the $j$-th
observation corresponding to the data set $d_j$, $j=1,\dots,N$, the joint
posterior is given by:
\begin{equation}
P(\df{220}, \dtau{220} | \{d_j\}) = P(\df{220}, \dtau{220}) \prod _{j=1}^N \frac{P(\df{220}, \dtau{220} | d_j) }{P(\df{220}, \dtau{220})}\,,
\end{equation}
where $P(\df{220}, \dtau{220})$ is the prior on $(\df{220},
\dtau{220})$. However, since here we assume the prior on $(\df{220},
\dtau{220})$ to be flat (or uniform), the joint posterior is equal to
the joint likelihood. We show these joint likelihoods on $(\df{220}, \dtau{220})$, as well as, the corresponding 1D marginalized distributions as filled grey curves in Fig.~\ref{fig:o1o2_events}. From the joint likelihood we obtain within 90\% credibility: 
\begin{equation}
\df{220}=0.02^{+0.04}_{-0.04}\,,\quad \dtau{220}=0.10^{+0.14}_{-0.14}\,.
\end{equation}

However, in most non-GR theories, the deviations parameters $(\df{220},\dtau{220})$ depend in general on the source's parameters, so if GR were to be wrong, one should expect their value to vary across the GW signals observed by LIGO and Virgo.
As described in Ref.~\cite{Abbott:2020jks}, we can relax the assumption of a constant deviation across all events by using the hierarchical inference technique originally proposed in Refs.~\cite{Zimmerman:2019wzo,Isi:2019asy}. The general idea behind this technique is to assume that the non-GR parameters $(\df{220},\dtau{220})$ are drawn from a common underlying distribution, whose properties can be inferred from the population of events. Following~\cite{Zimmerman:2019wzo,Isi:2019asy,Abbott:2020jks} we model the population distribution with a Gaussian  $\mathcal{N}(\mu,\sigma)$ of unknown mean $\mu$ and standard deviation $\sigma$. Under those assumptions, the goal is then to directly measure a posterior distribution $P(\mu, \sigma |  \{d_j\}) $ for $\mu$ and $\sigma$ from a joint analysis of all the GW events. If GR is correct, then this posterior should be consistent with $\mu=0$ and $\sigma=0$. From Bayes' theorem it follows that~\cite{Isi:2019asy}:
\begin{equation}
P(\mu, \sigma |  \{d_j\}) \propto P(\mu,\sigma)\prod _{j=1}^N P(d_j |\, \mu, \sigma) \,,
\end{equation}
where $P(\mu,\sigma)$ is the prior (also known as hyperprior) on ($\mu,\sigma$), and $P(d_j | \mu, \sigma) $ can be written in terms of the individual likelihoods of a given non-GR parameter $\xi_{\text{nGR}}$ using~\cite{Isi:2019asy}
\begin{equation}
P(d_j | \mu, \sigma) = \int P(d_j |\, \xi_{\text{nGR}})  P(\xi_{\text{nGR}} |\, \mu, \sigma) \,d\xi_{\text{nGR}} \,.
\end{equation}
Here $P(\xi_{\text{nGR}} | \,\mu, \sigma) = \mathcal{N}(\mu,\sigma)$ by construction and $P(d_j |\, \xi_{\text{nGR}})$ is the likelihood for the parameter $\xi_{\text{nGR}}$ for a given event $d_j$ that is computed from the standard parameter estimation analysis. After obtaining $P(\mu, \sigma |  \{d_j\})$, we can then infer a population distribution for a given non-GR parameter $\xi_{\text{nGR}}$ using~\cite{Isi:2019asy}:
\begin{equation}\label{hier_combine}
P(\xi_{\text{nGR}} | \{d_j\}) = \int P(\xi_{\text{nGR}} |\, \mu, \sigma)\,P(\mu, \sigma |  \{d_j\})\,d\mu\,d\sigma \,.
\end{equation}
Notice that, if we fix $\sigma=0$, this approach is equivalent to assume that all events share the same non-GR parameter $\xi_{\text{nGR}} =\mu$ and Eq.~\eqref{hier_combine} reduces to the joint likelihood~\cite{Zimmerman:2019wzo}. In practice, we use the \texttt{stan}-based code~\cite{stan} developed and used in Refs.~\cite{Isi:2019asy,Abbott:2020jks} to obtain $P(\mu, \sigma |  \{d_j\}) $ and compute $P(\xi_{\text{nGR}} | \{d_j\})$. We note that the current implementation of this analysis, as originally defined in~\cite{Isi:2019asy}, is only defined for 1D posteriors, therefore below we only show 1D posteriors for the hierarchical analysis.

The posteriors for $\df{220}$ and $ \dtau{220}$ obtained with this technique are shown in Fig.~\ref{fig:o1o2_events} (dash-dotted curves) with corresponding median and 90\% credible interval given by: 
\begin{equation}
\df{220}=0.03^{+0.10}_{-0.09}\,,\quad \dtau{220}=0.10^{+0.44}_{-0.39}\,.
\end{equation}
Compared to Ref.~\cite{Abbott:2020jks} these constraints are almost a factor $\sim 4$ more stringent for $\df{220}$ and a factor $\sim 2$ for $\dtau{220}$. Similar improvements hold for the hyperparameters: $\df{220}\,(\mu=0.03^{+0.06}_{-0.05},\,\sigma<0.09)$ and $\dtau{220}\,(\mu=0.11^{+0.21}_{-0.19},\,\sigma<0.39)$.

\begin{table*}

\begin{tabular}{l|c|c|c|c|c|c}
\toprule
Event & $\fngr{220}$ (Hz) & $\taungr{220}$ (ms) & $(1+z)M_f/\Mo$ & $\chi_f$ & $(1+z)M_{\text{f}}^{\rm IMR}/\Mo$ & $\chi_{\text{f}}^{\rm IMR}$ \\[0.075cm]
\midrule
\hline

GW150914 &
$257.6^{+17.0}_{-12.8}$ &
$4.49^{+1.09}_{-0.95}$ &
$71.0^{+8.7}_{-10.3}$ &
$0.77^{+0.09}_{-0.18}$ &
$67.3^{+2.7}_{-2.6}$ &
$0.67^{+0.03}_{-0.04}$
\\[0.075cm]

GW170104 &
$291.4^{+14.7}_{-30.1}$ &
$5.53^{+3.47}_{-2.40}$ &
$73.8^{+11.1}_{-19.8}$ &
$0.89^{+0.07}_{-0.36}$ &
$56.9^{+3.0}_{-3.0}$ &
$0.65^{+0.05}_{-0.07}$
\\[0.075cm]

GW190519\_153544 &
$123.6^{+11.9}_{-13.0}$ &
$10.33^{+3.56}_{-3.07}$ &
$155.5^{+24.0}_{-29.9}$ &
$0.81^{+0.10}_{-0.28}$ &
$144.1^{+14.5}_{-16.2}$ &
$0.78^{+0.08}_{-0.14}$
\\[0.075cm]

GW190521\_074359 &
$204.6^{+14.6}_{-11.7}$ &
$5.32^{+1.48}_{-1.21}$ &
$86.4^{+12.2}_{-14.3}$ &
$0.73^{+0.12}_{-0.26}$ &
$87.1^{+3.3}_{-3.8}$ &
$0.70^{+0.03}_{-0.05}$
\\[0.075cm]

GW190630\_185205 &
$247.8^{+31.8}_{-52.8}$ &
$3.87^{+2.37}_{-1.80}$ &
$65.6^{+18.8}_{-41.5}$ &
$0.62^{+0.27}_{-0.62}$ &
$66.2^{+4.0}_{-3.2}$ &
$0.70^{+0.05}_{-0.08}$
\\[0.075cm]

GW190828\_063405 &
$257.8^{+201.3}_{-27.8}$ &
$4.23^{+4.17}_{-1.92}$ &
$67.4^{+25.7}_{-30.0}$ &
$0.76^{+0.20}_{-0.73}$ &
$75.8^{+5.0}_{-5.0}$ &
$0.74^{+0.04}_{-0.06}$
\\[0.075cm]

GW190910\_112807 &
$174.2^{+11.7}_{-7.5}$ &
$9.52^{+3.13}_{-2.68}$ &
$123.5^{+14.7}_{-18.1}$ &
$0.90^{+0.05}_{-0.11}$ &
$94.9^{+7.6}_{-8.6}$ &
$0.72^{+0.08}_{-0.04}$
\\[0.075cm]

\bottomrule
\end{tabular}
\caption{The median and symmetric 90\% credible intervals of the remnant properties. The first two columns represent the frequency and damping time of the $(2,2)$ QNM measured using the $\pSEOB$ model. The next two columns are the mass and spin of the remnant object estimated from the complex QNM frequencies by inverting the fitting formula in~\cite{Berti:2005ys}. The last two columns represent final mass and spin estimates predicted using NR fitting formulae from a $\SEOB$ parameter estimation.}
\label{tab:qnm_o1o2_results}
\end{table*}



\section{Discussion}
\label{sec:discussion}
We have built a parameterized IMR waveform model, $\pSEOB$, that can 
measure the QNM complex frequencies of the remnant object formed through 
the merger of BHs with aligned or antialigned spins, thus extending  
previous work, which was limited to non-spinning BHs~\cite{Brito:2018rfr}. 
The $\pSEOB$ model was recently used to infer the QNMs of some of the 
BBH's remnants detected by LIGO and Virgo during O3a~\cite{Abbott:2020jks}. 

After testing our method to infer the QNM frequencies with GR and non-GR synthetic-signal 
injections in Gaussian noise, we have applied it to LIGO and Virgo real data. 
We have analyzed GW events in O1, O2 and O3a (a total of 4 new events) that were not examined in Ref.~\cite{Abbott:2020jks} 
with the $\pSEOB$ model (see Table~\ref{tab:qnm_o1o2_results}). After combining our new 
results with the other GW events in O3a investigated with our method~\cite{Abbott:2020jks}, 
we have obtained more stringent bounds on the dominant (or least-damped) QNM $(\ell=2,m=2)$.

More specifically, as expected, the single GW event providing the best constraint
on the least-damped QNM to date continues to be GW150914 $(\df{220}=0.05^{+0.11}_{-0.07}, 
\dtau{220}=0.07^{+0.26}_{-0.23})$, although the bounds on 
$\dtau{220}$ from GW150914 are comparable to GW190521\_074359, 
the second-most precisely measured event in our list.
In addition, in the most agnostic and
conservative scenario where we combine the information from different
events using a hierarchical approach~\cite{Zimmerman:2019wzo,Isi:2019asy}, we obtain 
at $90\%$ credibility $(\df{220}=0.03^{+0.10}_{-0.09}, \dtau{220}=0.10^{+0.44}_{-0.39})$. Thus, 
our results constrain the frequency (decay time) of the least damped QNM to be within $\sim
10\%$ ($\sim 40\%$) of the GR prediction --- an improvement of a factor of $\sim 4$ ($\sim 2$) 
over the results obtained with the $\pSEOB$ model in Ref.~\cite{Abbott:2020jks}.

Furthermore, when assuming that the deviations from GR do not vary appreciably over the GW events 
that we have analyzed, and combine the likelihood functions, we have obtained the most stringent bounds  
on the dominant QNM $(\df{220}=0.02^{+0.04}_{-0.04}, \dtau{220}=0.10^{+0.14}_{-0.14})$. Those 
constraints are compatible, but slightly better (especially for the decay time) than the ones recently reported in Ref.~\cite{Carullo:2021dui} 
(see first row in Table II therein), where QNM frequencies were inferred using only the post-merger part of 
the signal. We note that Ref.~\cite{Carullo:2021dui} used a larger set of GW events from O1, O2 and O3a than we did, although we expect 
that the extra GW events will not contribute significantly to the combined bound since they have lower SNRs.

These constraints could in principle be used to constrain specific
non-GR theories and exotic compact objects~\cite{Glampedakis:2017cgd,Cardoso:2019rvt,Maggio:2020jml}. 
However, QNM computations in non-GR theories have only been done in an
handful of cases, mostly focusing on non-rotating or slowly-rotating
BH solutions~\cite{Ferrari:2000ep,Molina:2010fb,Pani:2009wy,Blazquez-Salcedo:2016enn,Blazquez-Salcedo:2017txk,Brito:2018hjh,Franciolini:2018uyq,Cardoso:2018ptl,Tattersall:2018nve,Tattersall:2019nmh,Blazquez-Salcedo:2019nwd,Silva:2019scu,Glampedakis:2019dqh,Blazquez-Salcedo:2020jee,Blazquez-Salcedo:2020caw,Cano:2020cao,Wagle:2021tam,Pierini:2021jxd}, 
or relying on the eikonal/geometric optics approximation to obtain
estimates of the QNMs for spinning BHs~\cite{Blazquez-Salcedo:2016enn,Glampedakis:2017dvb,Jai-akson:2017ldo}. The
only exceptions to this rule, that we are aware of, are computations
of the QNM of Kerr-Newman BHs in Einstein-Maxwell
theory~\cite{Pani:2013ija,Pani:2013wsa,Mark:2014aja,Dias:2015wqa} or estimates of the BBH in non-GR theories obtained
through a limited number of NR simulations~\cite{Okounkova:2019dfo,Okounkova:2019zjf}. Given these
limitations, our ability of going beyond a null test of GR and use our
results to impose precise constraints on non-GR theories with QNM
measurements is currently quite limited.

Despite these theoretical limitations there has been some recent effort to develop
parametrizations that could help mapping measurements of the
parameters $(\delta f_{\ell m 0}, \delta \tau_{\ell m0})$ onto
constraints to specific non-GR theories. This includes for example
the parametrization proposed in Ref.~\cite{Maselli:2019mjd}, recently
applied to LIGO-Virgo GW events in Ref.~\cite{Carullo:2021dui}, where
deviations from the GR QNMs explicitly depend on a perturbative
expansion in the BH spin and possible extra non-GR parameters, or the
proposal of Refs.~\cite{Cardoso:2019mqo,McManus:2019ulj} where
deviations from the GR QNMs are mapped onto generic small
modifications of the perturbation equations describing the QNMs. Other
examples also include proposals to map deviations from the GR QNMs to
coefficients in generic effective--field-theory
actions~\cite{Cardoso:2018ptl,Franciolini:2018uyq,Cano:2020cao} or to
directly relate measurements of the QNM complex frequencies to a
parametrized non-GR BH metric~\cite{Glampedakis:2017dvb,Suvorov:2021amy,Volkel:2020daa}, 
which could be then used jointly with measurements from the Event
Horizon Telescope to obtain stronger constraints on deviations from
GR~\cite{Volkel:2020daa,Volkel:2020xlc,Psaltis:2020lvx,Yang:2021zqy}.
We should note, however, that all these parametrizations are either
limited to non-spinning BHs or make use of a series expansion in the
BH spin which might limit their accuracy for highly-spinning BHs, unless 
the sensitivity of GW detectors will not allow us to access the higher 
coefficients in the spin series.

In the future, it would be important to test whether the $\pSEOB$  model
could be used to detect deviations in waveforms obtained through NR 
simulations of specific non-GR theories. Such results are still at their 
infancy and have so far only been done for a handful of theories, focusing mostly on proof-of-concept
simulations~\cite{Healy:2011ef,Berti:2013gfa,Cao:2013osa,Okounkova:2017yby,Hirschmann:2017psw,Witek:2018dmd,Okounkova:2019dfo,Okounkova:2019zjf,Okounkova:2020rqw,East:2020hgw}. Nonetheless,
given the recent efforts put forward in order to simulate BBHs in
non-GR theories, we hope that accurate non-GR IMR waveforms will become available in the near future.

Finally, an obvious generalization of this work would be to extend the
parameterized $\pSEOB$ model to generic precessing BBHs, which can in
principle be easily done using the recently developed multipolar EOB
model reported in Ref.~\cite{Ossokine:2020kjp}. Lastly, it will be relevant 
to include GR deviations in the $\pSEOB$ model also during the late inspiral 
and plunge stages of the BBH coalescence. On the other hand, GR deviations (notably 
deviations from PN theory) for the long inspiral stage are currently 
available in the $\pSEOB$ model and have been used to set bounds on the 
PN parameters in the GW phasing using LIGO and Virgo observations~\cite{Abbott:2018lct,
LIGOScientific:2019fpa,Abbott:2020jks}.


\section*{Acknowledgements}
\label{sec:acknowledgements}
The authors thank Max Isi for generously allowing us to use the code developed in Ref.~\cite{Isi:2019asy} for the hierarchical inference analysis, and Gregorio Carullo for carefully reading the manuscript and providing useful comments. The authors would like to thank everyone at the frontline of the Covid-19 pandemic. The authors are grateful for computational resources provided by the LIGO Laboratory and supported by the National Science Foundation Grants PHY-0757058 and PHY-0823459, as well as computational resources at the AEI, specifically the Hypatia cluster. R.B. acknowledges financial support from the European Union's Horizon 2020 research and innovation programme under the Marie Sk\l odowska-Curie grant agreement No. 792862, from the European Union's H2020 ERC, Starting Grant agreement no.~DarkGRA--757480, and from the MIUR PRIN and FARE programmes (GW-NEXT, CUP:~B84I20000100001). We also acknowledge support from the Amaldi Research Center funded by the MIUR program ``Dipartimento di Eccellenza'' (CUP:~B81I18001170001).

\appendix
\section{Study of systematics in ringdown measurements in real, non-Gaussian noise}\label{sec:noise_systematics}

Inferences of all parameters in this paper have been done under the
assumption that the noise in the detectors is stationary and
Gaussian. In other words, detector noise follows a normal distribution
with zero mean and a PSD, $S_n(f)$, that is not a function of time, at
least during the duration of the GW signal. This
allows us to write the Bayesian likelihood function in the form given
in Eqs.~(\ref{eq:likelihood}) and ~(\ref{eq:nwip}), and perform all the
parameter estimation that follows in the results sections. However,
LIGO-Virgo noise can often have features that deviate from
stationarity and Gaussianity. If such features are not taken into
account appropriately, final estimates of parameters can get
biased. Here we demonstrate one such case by injecting in real noise a
GW190521-like signal and showing how parameter estimates can be biased
when our description of detector noise is not complete.

\begin{figure}
\begin{center}
        \includegraphics[width=0.4\textwidth]{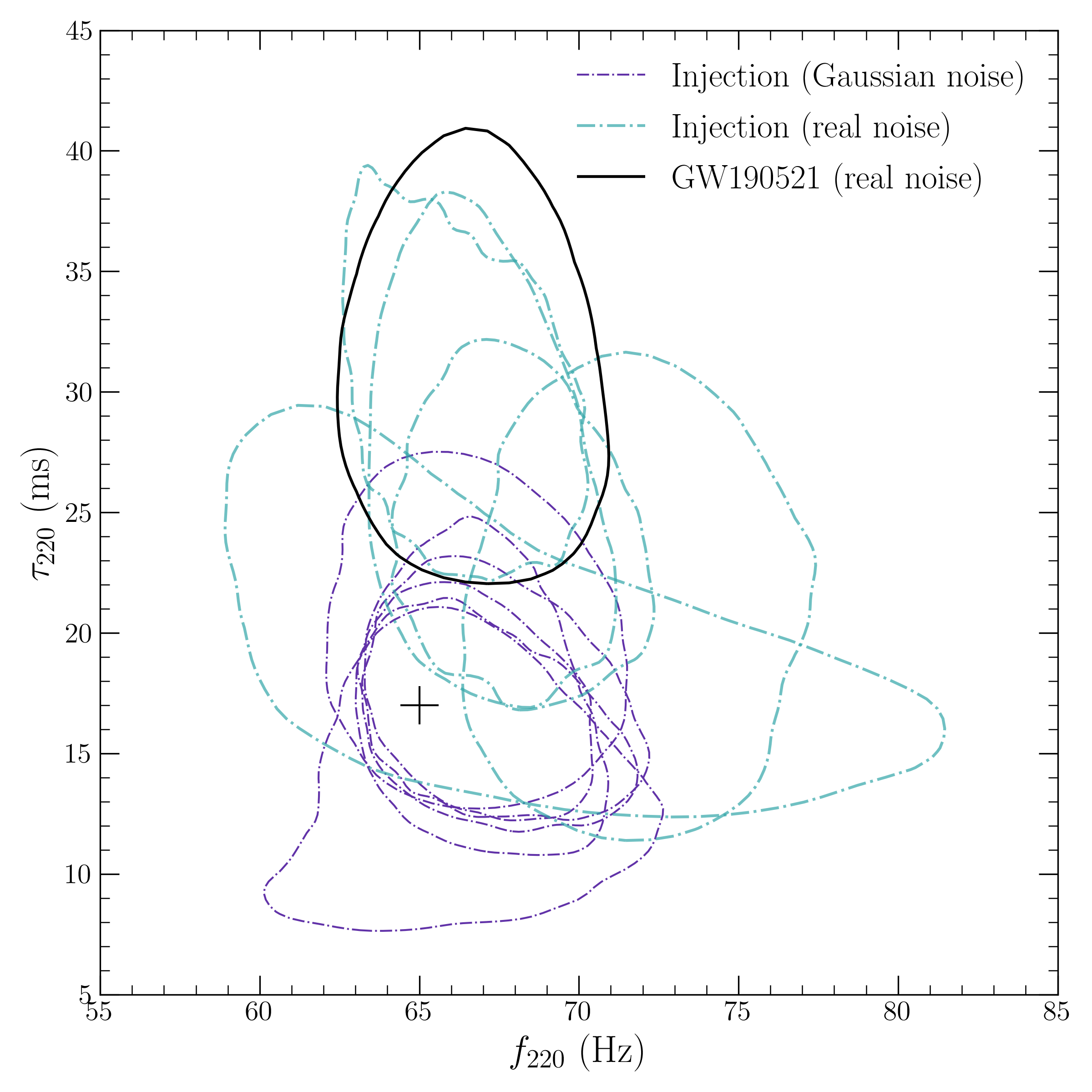}
        \caption{90 \% credible level on the posterior probability distribution of the frequency and damping time of $(2, 2)$ mode, $(\fgr{220}, \taugr{220})$ using synthetic \texttt{NRSur} signals with parameters similar to the GW event, GW190521, in Gaussian noise (grey dot-dashed lines) and real interferometric noise (green dot dashed lines). The GR prediction for the frequency and damping time is indicated by the black cross. While the Gaussian noise simulations are consistent with the prediction, at least 3 of the 5 real noise simulation are not. The black curve corresponds to the measurements of the real event GW190521 reported in Ref.~\cite{Abbott:2020jks}. All signals are recovered using the $\pSEOB$ model.}
        \label{fig:21g_systematics}
\end{center}
\end{figure}

We choose a spinning, precessing NR-surrogate model \texttt{NRSur}~\footnote{This waveform model is 
called \texttt{NRSur7dq4} in LAL.} (valid up to mass ratio 4) to simulate the actual GW190521 signal
observed by the LIGO and Virgo detectors~\cite{Abbott:2020tfl} (see
Table I of Ref.~\cite{Abbott:2020tfl})). The choice of the
\texttt{NRSur} model is motivated by the fact that it is the most
accurate model in the parameter range described by GW190521, because 
it is built by directly interpolating NR waveforms. In
Fig.~\ref{fig:21g_systematics}, we indicate with a black cross what
the injected \texttt{NRSur} signal predicts for the QNM $(\ell=2,m=2)$
frequency and damping time. For comparison, we also show with a
black solid curve the results obtained when recovering the actual
signal GW190521 with the $\pSEOB$ model. As seen in the plot,
while the measurement of the frequency is consistent with the
prediction, we overestimate the damping time.

To understand such offset in the decay time, we proceed as follows.
The actual GW190521 event was observed at a GPS time, 1242442967.61
seconds (roughly 03:02:49 UTC, May 21, 2019). We select a time period
of about 2.5 hours around this GPS time, create synthetic signals 
with the \texttt{NRSur} model and inject them in different stretches
of the real detector noise around the time of the actual GW event. The
PSDs of GW detectors are expected to vary over longer durations of
time, and hence the 2.5 hour stretch of noise we consider can be
assumed to have noise-properties similar to the time of the actual
event. Then, we perform Bayesian analysis against those injections 
using the $\pSEOB$ model. The results are indicated by green curves in
Fig.~\ref{fig:21g_systematics}. As it can be seen from the figure, for
3 of the 5 noise realizations, corresponding to $t_0-1$ hour,
$t_0+0.5$ hours, and $t_0+1$ hour, we recover a damping time similar to
the one obtained when using the $\pSEOB$ model against the actual event GW190521 
(black curve), where $t_0$ is the GPS time of the actual event. For
the other two noise realizations, the $\pSEOB$ model estimates 
consistently the damping time, but has an off-set frequency, 
while the fifth noise realization is consistent with both predictions.  
This study suggests that a bias in the measurements of the damping time 
for the actual event GW190521 can be explained as due to an incomplete 
description of the noise at the time of the event.

The reader might question the judiciousness of using an aligned-spin
waveform model, like $\pSEOB$, to measure a signal like GW190521 which
appears to be precessing, especially because an incomplete
understanding of the underlying signal can also lead to biases in
measured quantities, as we have already demonstrated in
Sec.~\ref{ssec:ngr_signal}. In order to explore possible effects of
missing information about in-plane spins in the $\pSEOB$ model, we repeat the
above study of injecting synthetic signals using \texttt{NRSur}
and recovering using the $\pSEOB$ model, but this time, instead of
using real detector noise, we use Gaussian noise (i.e., realizations
of noise sampled from a predicted detector PSD). Since the properties
of the noise are completely understood in this case, any residual
measurement biases can be completely attributed to diffferences in the
waveform model. The 2D posterior distributions of the frequency and
damping time measured using these Gaussian-noise signals are shown by
the grey curves in Fig.~\ref{fig:21g_systematics}. We find the
measurements to be completely consistent with the predictions of the
frequency and damping time, thus concluding that a lack of in-plane
spins in the $\pSEOB$ model does not affect our measurements of the
QNM properties. The fact that the measurement of ringdown quantities
are robust against an incomplete description of the inspiral signal 
is a crucial property of our method.

\section{Correlation of the binary's total mass with the non-GR parameters}
\label{sec:correlation}

\begin{figure*}
\begin{center}
        \includegraphics[width=0.45\textwidth]{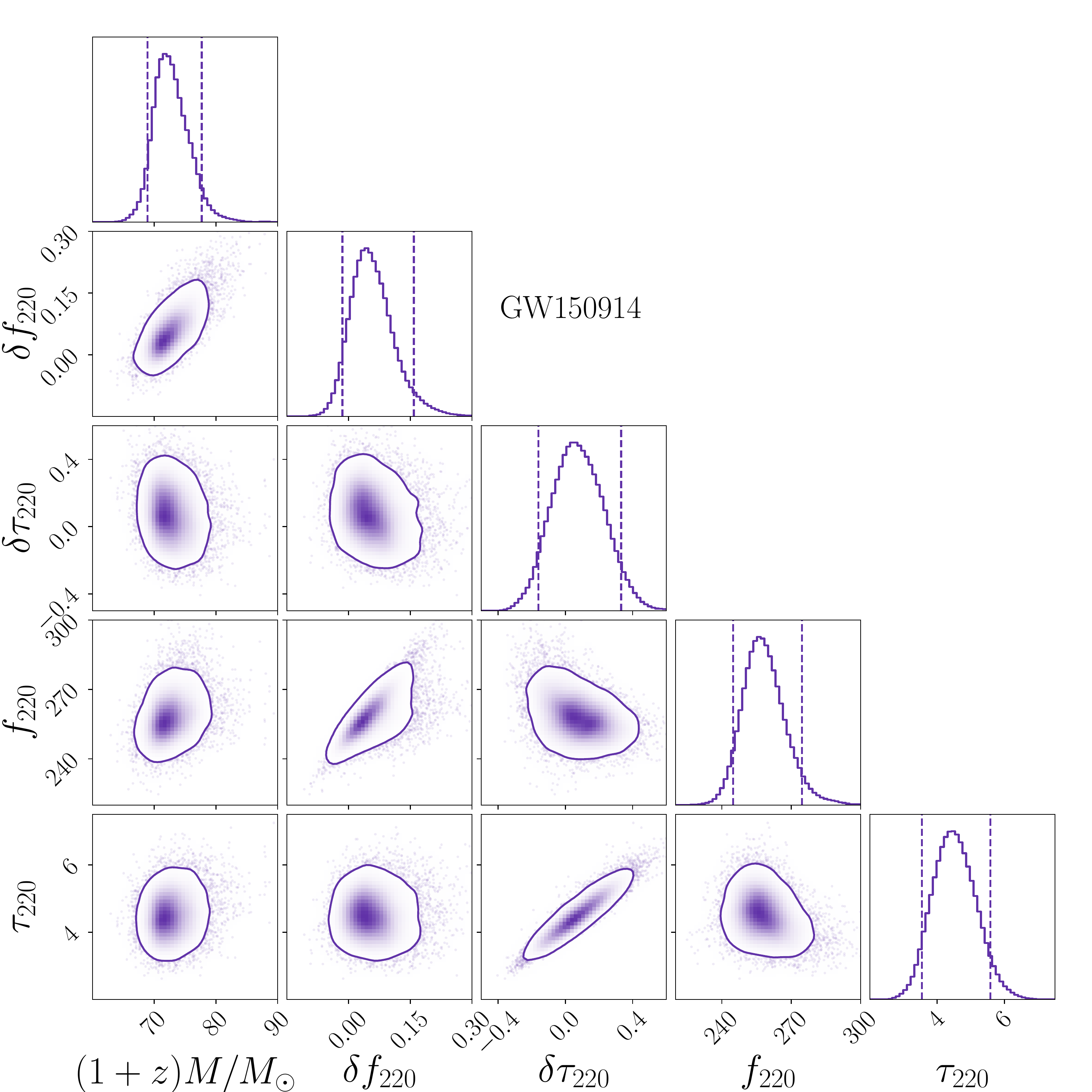}\includegraphics[width=0.45\textwidth]{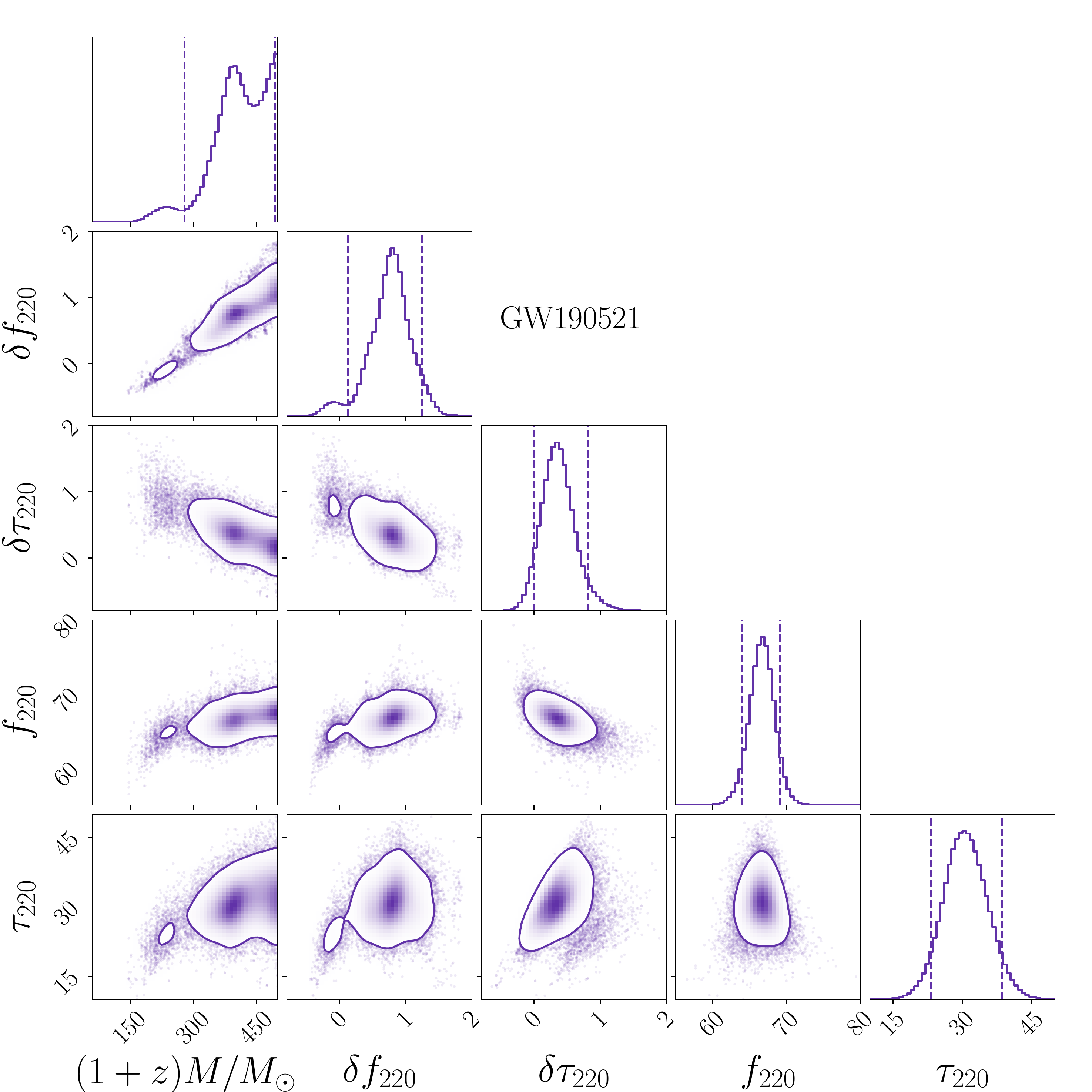}
        \caption{Corner plots showing the correlations between the detector-frame total $(1+z)M$, the non-GR deviations $(\df{220}, \dtau{220})$ and the reconstructed frequency and damping time $(f_{220},\tau _{220})$. The left panel shows results for GW150914, for which ${\rm SNR}> 8$ in both the pre- and
post-merger phase of the signal, and the correlations are absent. The right panel shows the results for GW190521,  which has ${\rm SNR}> 8$ in the post-merger phase, but not in the pre-merger phase, and the correlations are present.}
        \label{fig:correlations}
\end{center}
\end{figure*}

As mentioned in the main text, for low-SNR events with negligible
higher-modes and for which only the post-merger is detectable, there
is a strong degeneracy between the binary's total mass and the non-GR
deviations $(\df{220}, \dtau{220})$. For those cases, only the
reconstructed frequency and damping time $(f_{220},\tau _{220})$ can
be independently measured from the data. To justify this statement, 
in Fig.~\ref{fig:correlations} we show
corner plots that illustrate the correlations between the non-GR
parameters $(\df{220}, \dtau{220})$, the detector-frame total mass
$M(1+z)$ and the reconstructed frequency and damping time
$(f_{220},\tau _{220})$ for GW150914 (left panel), corresponding to an
event for which both the pre- and post-merger phase are measurable,
and for GW190521 (right panel), an event in which the post-merger has
${\rm SNR}> 8$, but the pre-merger has an SNR below 8. For GW190521,
due the strong degeneracy between $\df{220}$ and $M(1+z)$, the 1D
posterior for $M(1+z)$ is pushed towards the upper boundary of its
prior, despite the very wide prior employed in the analysis. This in
turn renders the measurement of $\df{220}$ (and to a lesser degree of
$\dtau{220}$) highly dependent on the upper boundary of the total mass
prior. On the other hand, this issue does not significantly affect the
posteriors for the reconstructed quantities $(f_{220},\tau _{220})$
which are well measured and nearly independent on the upper prior
boundary for $M(1+z)$. This is to be contrasted with the results for GW150914. In this case, 
the extra information coming from the pre-merger phase allows to break 
the degeneracy between the non-GR parameters and the total mass, 
and therefore both $(\df{220}, \dtau{220})$ and $M(1+z)$ can 
be measured at the same time. 

%
\bibliographystyle{apsrev}
\bibliography{intro_paper}

\end{document}